\documentclass[aps,floatfix,prd,showpacs]{revtex4}
\usepackage{graphicx}
\usepackage{dcolumn}
\usepackage{bm}
\usepackage{float}
\usepackage{color}

\begin{document}

\title{Inflation in Loop Quantum Cosmology}
\author{Anshuman Bhardwaj$^{1,2}$, Edmund J. Copeland$^{1}$, Jorma Louko$^{3}$}
\affiliation{
$^{1}$School of Physics and Astronomy, University of Nottingham, Nottingham NG7 2RD, United Kingdom\\
$^{2}$Department of Physics and Astronomy, Louisiana State University, Baton Rouge, LA 70803, USA\\
$^{3}$School of Mathematical Sciences, University of Nottingham, Nottingham NG7 2RD, United Kingdom}
\date{Revised March 2019}

\begin{abstract}
We develop a consistent analytic approach to determine the conditions under which slow roll inflation can arise when the inflaton is the same scalar field that is responsible for the bounce in Loop Quantum Cosmology (LQC). We find that the requirement that the energy density of the field is fixed at the bounce having to match a critical density has important consequences for the future evolution of the field. Initially we consider the case of a generic potential which has a minimum, and we find different scenarios depending on the initial velocity of the field and whether it begins life in a kinetic energy or potential energy dominated part of its energy density. For chaotic potentials that start in a kinetic dominated regime, we find an initial phase of superinflation independent of the shape of the potential followed by a damping phase which slows the inflaton down, forcing it to turnaround and naturally enter a phase of slow-roll inflation. If we begin in a potential energy dominated regime, then the inflaton undergoes a period where the corrections present in LQC damp its evolution once again forcing the field to turnaround and enter a phase of slow-roll inflation. On the other hand we show for Starobinsky inflation that inflation never occurs when we begin in a potential energy dominated regime. What we would normally call potential dominated in traditional Starobinsky inflation where the field lives in its plateau regime, for the case of LQC this corresponds to being in a kinetic energy dominated regime. The requirement that damping slows the field down sufficiently to enter slow roll places tighter constraints on the initial value of the field for successful inflation than in the conventional case. Comparing our analytic results to the published numerical ones, we find remarkable agreement especially when we consider the different epochs that are involved. In particular the values of key observables obtained from our analytical and the published numerical solutions are in excellent agreement, opening up the possibility of using our results to obtain analytic results for the evolution of the density perturbations in these models. 
\end{abstract}

\maketitle

\section{Introduction}
\noindent The Inflationary paradigm \cite{Inflation} is regarded by many (but not everyone \cite{Steinhardt:2011zza}) as providing the most promising description of the early universe, in particular through the generation of quantum fluctuations in the inflaton field which by coupling to the spatial curvature of the universe act as seeds for the observed anisotropies in the cosmic microwave background (CMB) \cite{Ade:2015xua} and the large scale structure of our universe \cite{Tegmark:2003ud}. Usually based on the assumption of quantum fields propagating on a  classical spacetime described by General Relativity (GR), the current formulation fails to account for how inflation may arise when we want to extend the analysis into regimes where we would like to resolve the primordial singularity \cite{Ashtekar:2016wpi}. This of course is tricky territory and the best we have to go on at the moment are models describing such a regime, models which themselves contain a number of assumptions. In this paper, we will look at inflation arising in a class of such models, bouncing cosmologies, and ask the question how do the conditions associated with the bounce affect the background evolution of the inflaton field compared to the usual situation without a bounce.  \\~\\
There are a number of approaches to a quantum theory of gravity, of which some include bouncing regimes. Of many tasks they have to satisfy, two include resolving the initial singularity, whilst also allowing slow-roll inflation to naturally arise (if indeed it does). Various cosmological inflation models have been proposed in the context of String/M-theory\cite{Burgess,Kachru,Conlon:2005jm,McAllister:2007bg,Baumann:2014nda} (see also \cite{Obied:2018sgi,Agrawal:2018own} for recent discussions on the possible problems facing inflation model building in the context of the string swampland), but far fewer examples exist showing how it can be embedded in say Loop Quantum Gravity (LQG)\cite{LQG}. The latter is a background independent theory which uses polymer techniques to quantize the Hamiltonian of GR but written in terms of holonomy and triad variables. This gives rise to the Gauss, Diffeomorphism and Hamiltonian constraint operators which are quantum analogues of Einstein's equations. A reduction of LQG known as Loop Quantum Cosmology(LQC)\cite{Ashtekar-Singh,Ashtekar-Bojowald-Lewandowski,Ashtekar:2016ecx} is obtained by reducing GR to its cosmological degrees of freedom before quantising with LQG techniques - in some sense it is akin to the mini-superspace analogue of LQG. The simplicity of the equations associated with LQC has led to a great deal of work in the subject - for a couple of nice recent discussions see \cite{Ashtekar:2016wpi,Ashtekar:2016ecx}. The Hamiltonian constraint associated with LQC can be thought of as describing the quantum evolution of a `dressed metric' for the background cosmology associated with the early universe. An effective or semi-classical description was developed in \cite{Ashtekar-Pawlowski-Singh}, which showed how a connection could be made with the standard classical cosmological evolution equations, the change being that the effective Friedmann equation contains an additional negative $\rho^{2}$ term (where $\rho$ is the total energy density of any matter present in the Universe), which causes the contracting universe to undergo a Big Bounce (at $t=0$) as soon as a critical density of matter is reached. At the level of the perturbations in the system, the dressed metric approach has been developed to describe the evolution of perturbations on LQC spacetimes \cite{Agullo-Ashtekar}. If the background state is sharply peaked and there is no quantum backreaction of the perturbations on the background then a modified Mukhanov-Sasaki equation including quantum corrections is obtained. \\~\\
To summarise, in LQC, and in more general bouncing cosmologies, we have an effective Friedmann and Klein-Gordon equation subject to the constraint that the energy density at the bounce is fixed. These equations describe the journey of an inflaton field in the effective LQC spacetime, which has been studied numerically for the cases of quadratic \cite{Ashtekar-Sloan} and Starobinsky\cite{Bonga-Gupt} potentials for the inflaton field, as well as for the case of power law potentials of the form $\phi^n$ where $n<2$ \cite{Shahalam:2017wba}. Recently there has been interest in understanding the nature of inflation in a class of modified LQC models which emerge by keeping the Lorentzian term explicit in the Hamiltonian constraint \cite{Li:2018opr,Li:2018fco}. \\~\\
There is an important aspect of our assumptions that should be briefly discussed. As we shall shortly see, the precise solutions depend on a constant number known as the Barbero-Immirzi parameter, whose precise value is a matter for some debate (see the review of Perez \cite{Perez:2017cmj} for details). In particular it is crucial to the existence of the particular bounce solution we will discuss in the paper. However, there are alternative approaches to LQG in which the assumption can be removed. For example in \cite{Wang:2018bdg} this is done by using a conformal or scaling symmetry and leads to a new loop quantization arising from a conformally generalized Holst action principle with no Barbero-Immirzi ambiguity. In the context of our work, the result is interesting because in the case of these scaling invariant cosmologies, inflation can naturally occur, for example see \cite{Ferreira:2018qss}.
\\~\\
In this paper, within the context of LQC, we consider analytically the evolution of the inflaton field from the initial bounce in a number of interesting potentials, comparing our findings where appropriate to the published numerical results of Refs.~\cite{Ashtekar-Sloan} and \cite{Bonga-Gupt}. We  find that the specific evolution depends on which of the terms are dominating the energy density as we emerge from the bounce. In particular we develop analytic techniques to describe the field's evolution and approach to slow roll for situations where the initial inflaton energy density is dominated by either its potential or kinetic terms. What makes things a bit different in the case of these bouncing models is that there is a constraint on the total energy density of the inflaton field at the  bounce, and this has an impact on its subsequent evolution. Our goal is to describe the circumstances under which slow roll inflation takes place for these different initial conditions including the number of e-folds we can expect as a function of the potentials we consider. Comparisons with published numerical results of Refs~\cite{Ashtekar-Sloan} and \cite{Bonga-Gupt}, will show how our analytic solutions provide excellent approximations to their numerical solutions.  \\~\\
The rest of the paper is as follows: in section~\ref{framework} we establish the basic framework used to consider the evolution of the universe  including a scalar field through a LQC bounce. In section~\ref{phases} we consider the case of a generic inflaton potential which admits slow roll solutions, and we develop analytic solutions to describe the evolution of the inflaton field in situations which are either initially kinetic energy (KE) or potential energy (PE) dominated at the bounce. In the former we show the onset of a superinflation regime as we leave the bounce and the field evolves up the scalar field potential, followed by damping, turnaround (where the potential energy is dominating) and finally entry into the slow roll regime as the field begins to roll back down the potential. In the latter, we see an initial period of damping where the quantum gravity induced terms slow the field down into a slow roll inflating epoch. In both situations, we will see that LQC or a similar bouncing model where the energy density is bounded at the bounce provides a viable explanation for the pre-inflationary regime and a smooth transition to the standard inflationary regime. In section~\ref{comparison} we reproduce the numerical results of \cite{Ashtekar-Sloan,Bonga-Gupt} and compare them with our analytic solutions for three classes of inflationary potentials, a quadratic chaotic potential, the Starobinsky potential (associated with $R^2$ inflation) and the quartic chaotic model. In each case we identify the epoch when the different regimes are reached and provide details of the field evolution at those epochs, showing how well the approximations work. Finally we conclude in section~\ref{conc}.   \\~\\
A note on conventions: we will work in the units $\hbar=c=1$. We explicitly keep $m_{\rm Pl}$ in the equations. Also we will assume that there is no cosmological constant and that the universe is spatially flat i.e. $\Lambda=0=k$. A subscript 'B' denotes at the bounce. 

\section{The Framework}
\label{framework}
\noindent We assume a spatially flat Friedmann Robertson Walker spacetime with metric 
\begin{equation}
ds^2 = - dt^2 + a^2(t)(dx^2 + dy^2 + dz^2)
\end{equation} 
where $a(t)$ is the scale factor. Now in LQC the key equations that describe the evolution of the scalar field $\phi(t)$ and $a(t)$ are the effective Friedman equation with holonomy corrections and the Klein Gordon  equation \cite{Ashtekar-Pawlowski-Singh}\\
\begin{equation} \label{Friedmann}
	H^{2} = \frac{8\pi}{3m^2_{\rm Pl}}\rho \Big(1-\frac{\rho}{\rho_{c}} \Big)
\end{equation}

\begin{equation} \label{KG}
	\ddot{\phi}+3H\dot{\phi}+V'(\phi) = 0
\end{equation}
where $H = \dot{a}/a$ is the Hubble parameter ($\dot{a} \equiv da/dt$), $\rho=\frac{\dot{\phi}^{2}}{2}+V(\phi)$ is the total energy density of what will become the inflaton field with potential $V(\phi)$ and $V'(\phi) \equiv dV/d\phi$. We will be considering the case where this is the only contribution to the energy density, which is justified in the regimes where we expect inflation to occur. The energy density satisfies the constraint $0 \leq \rho \leq \rho_{c}$ where $\rho_{c}= \frac{\sqrt{3}}{32\pi^{2}\gamma^{3}}m_{\rm Pl}^{4} \approx 0.41 m_{\rm Pl}^{4}$ is the critical density and $\gamma\approx0.2375$ is the Barbero-Immirzi parameter. 

Before we go on to discuss the physics of these equations, a few words of justification for their use is in order. In particular the dramatic modification of the usual Friedmann equation as seen in equation~(\ref{Friedmann}), where an extra term has appeared on the right hand side involving the square of the energy density. It arises from the way in which quantization occurs in LQG. The theory uses polymer techniques to quantize the Hamiltonian of GR but it is written in terms of holonomy and triad variables. This is in contrast to the usual Wheeler DeWitt quantization approach (see \cite{LQG,Kiefer:2004gr} for details). In terms of the holonomy and triad variables, new Gauss, Diffeomorphism and Hamiltonian constraint operators are obtained which are quantum analogues of Einstein's equations, and we can think of this as the evolution equations associated with a quantum geometry. LQC emerges as a truncation of the full LQG, by reducing GR to its cosmological degrees of freedom before quantising with the holonomy techniques of LQG (see \cite{Ashtekar-Singh,Ashtekar-Bojowald-Lewandowski,Ashtekar:2016ecx} for details). In particular the associated Hamiltonian constraint can be interpreted as the quantum evolution of a `dressed metric' for the background cosmology associated with the early universe. An effective or semi-classical description was developed in \cite{Ashtekar-Pawlowski-Singh} and in it they showed that a connection could be made with the standard classical Friedmann equation by effectively moving the quantum geometry effect to the right side of the equation where it manifests itself as an additional negative $\rho^{2}$ term. The implication for cosmology is huge, especially in the high curvature (early universe regime) as it causes the contracting universe to undergo a Big Bounce (at $t=0$) as soon as a critical density of matter is reached.

A second aspect that we should discuss is the particular value we have associated with the Barbero-Immirzi parameter, namely $\gamma\approx0.2375$. The chosen value was initially discussed by Meissner in \cite{Meissner:2004ju}. It was chosen to  match a state-counting calculation of the black hole entropy to the Bekenstein-Hawking entropy and was used in the work of Refs. \cite{Ashtekar-Sloan,Bonga-Gupt} which are the results we are comparing our work to. However, we should point out that 
since Meissner's calculation, the LQG community have derived new state-sum constructions, in which the Bekenstein-Hawking entropy can be obtained by any value of the Barbero-Immirzi parameter (for a discussion see section IV G of Perez's review \cite{Perez:2017cmj}). In that case, there is no distinguished real value that emerges from these new state-sum constructions,  as opposed to mathematical simplicity arguments for the imaginary values $\pm i$. We will not be attempting to go into the regime of imaginary Barbero-Immirzi parameters here. Relaxing the value of the parameter from that of Meissner does not seem to have received systematic attention in the LQC literature, and we will not embark on a systematic quantitative analysis of this question, however, as the results depend smoothly on the parameter in at least some neighbourhood of the Meissner value, the results will not undergo qualitative changes under small real changes in the parameter.

At the bounce, the density is maximum $\rho=\rho_{c}$ ($H=0$) and from there on it monotonically decreases with time. This implies that the values of $\phi$ and $\dot{\phi}$ will be initially bounded at the bounce, and we can find ourselves in (i) Kinetic dominated, (ii) Potential dominated or of course (iii) in a regime where the energy is equally distributed between the two. We will concentrate here on the case of either Kinetic or Potential energy domination initially. It turns out that these regimes provide an excellent approximation even when the two contributions are of the same order. For each case there are two more possibilities, the initial velocity of the field can be positive or negative sending the field to either larger or smaller values of $\phi$.   \\~\\  
 Thus there are four ways in which the inflaton can move on a potential depending on its initial conditions and the symmetries of the potential. We will be describing them in detail but here we give a brief overview of the key features. Recall, a subscript $B$ denotes the value at the bounce, and we are considering the case of symmetric chaotic models (such as $m^2 \phi^2, \lambda \phi^4$) unless otherwise stated. 
\begin{enumerate}
\item $KE_{B}\gg PE_{B}$ with $\phi_B>0$ or $\phi_B<0$ and $\dot{\phi}_{B}>0$ : the inflaton begins life in a KE dominated regime and we see that superinflation occurs as $H$ increases from zero to a maximum value and the scale factor increases as $a(t)\approx(1+4\pi\rho_{c}t^{2}/m^2_{\rm Pl})$ during this period. The phase draws to a close as the PE becomes comparable to the KE eventually matching it. As the field continues slowing down, we enter a PE dominated regime in which the inflaton slows down as it is damped eventually reaching the turnaround point ($\dot{\phi}=0$). It reverses direction and soon enters the standard slow-roll regime from which the observed density fluctuations emerge. The symmetry of the situation means it also applies to the case where $\phi_B<0$ or $\phi_B>0$ and $\dot{\phi}_{B}<0$. It can also occur in the Starobinsky model, for $\phi_B<0$ and $\dot{\phi}_{B}>0$. We see this type of behaviour in Tables~\ref{Table1}, \ref{Table3} and \ref{Table5}.
\item $KE_{B}\gg PE_{B}$ with $\phi_B >0$ and $\dot{\phi}_{B}<0$ : same as the case above but now the inflaton starts life moving in the opposite direction. It will again go through the superinflation regime, gradually slow down and enter a PE dominated regime but in contrast to the previous case it will now lead to slow-roll without the inflaton turning around. In particular we will see this occurring in Starobinsky inflation in Table~\ref{Table4}. This situation also applies to the Chaotic potential case where $\phi_B<0$ and $\dot{\phi}_{B}>0$ as seen in Tables~\ref{Table2} and \ref{Table6}.
\item $PE_{B}\gg KE_{B}$ with $\phi_B >0$  and $\dot{\phi}_{B}>0$ : the inflaton starts life in a PE dominated regime and begins to slow down as its PE grows. Eventually it stops and turns around entering a slow-roll regime. This situation also applies to the case where $\phi_B<0$ and $\dot{\phi}_{B}<0$. It is noteworthy that we do not see this regime in the Starobinsky model as can seen in Table~\ref{Table3}. 
\item $PE_{B}\gg KE_{B}$ with $\phi_B >0$ and $\dot{\phi}_{B}<0$  : the inflaton again starts life in a PE dominated regime but with its velocity in the opposite direction. Depending on the magnitude of $\dot{\phi}_B$, we enter a period of ordinary chaotic inflation, and the field enters a slow-roll phase without the need of turning around. This situation also applies to the case where $\phi_B<0$ and $\dot{\phi}_{B}>0$. Once again we will see that this cannot occur in the Starobinsky model as seen in Table~\ref{Table4}. Inflation in that case always emerges out of the KE dominated regime. 
\end{enumerate}
Before we conclude this section we would like to discuss the pivot point which allows us to connect theory with observations. Whether solving the system numerically or analytically, knowing how the fluctuations behave on a particular length scale where an observation is made allows us to normalise the theory and thereby determine an underlying parameter of the model such as a mass scale or self coupling. The largest observable mode of the CMB is $k_{*}=0.002Mpc^{-1}$
and this in turn gives the value of the amplitude of scalar perturbations as $A_{s}=(2.474\pm0.116)\times10^{-9}$ and the spectral index $n_{s}=0.9645\pm0.0062$ \cite{Ade:2015xua,Ade:2015lrj}. In potential driven slow roll inflation, the slow roll parameter is given by  $\epsilon_{V} = \frac{m^2_{\rm Pl}}{16\pi}\left(\frac{V'}{V}\right)^{2}$ (which is related to the Hubble slow roll parameter $\epsilon_{H} = -\frac{\dot{H}}{H^{2}}$). The number of e-folds between times $t_i$ and $t_f$ are given in terms of the scale factor $a(t)$ as  $N= \ln\frac{a(t_{f})}{a(t_{i})}= \int_{t_{i}}^{t_{f}} Hdt \simeq -\frac{8 \pi}{m^2_{\rm Pl}} \int_{\phi_i}^{\phi_f} \left(\frac{V(\phi)}{V'(\phi)}\right) d\phi$. Using these values and the number of e-folds $N_{*}$ from the time when the $k_{*}$ mode exited the horizon to the end of slow-roll inflation defined by $\epsilon_V=1$, one can determine the value of say the mass parameter in the quadratic chaotic potential and the observables $(\epsilon_{V*},H_{*},\phi_{*},\dot{\phi}_{*})$ at the time when the pivot mode exited the horizon (58 e-foldings before the end of inflation) as is discussed in \cite{Bonga-Gupt}. As an example for the case of the quadratic potential $V(\phi) = \frac{1}{2} m^2 \phi^2$, we find $\phi_{*}=\pm3.15 m_{\rm Pl}$ and $m=1.21\times10^{-6} m_{\rm Pl}$.

\section{The Different Phases of Evolution}
\label{phases}
\noindent We now begin a detailed analysis of the dynamics associated with the various scenarios outlined in section~\ref{framework}. Initially this will be in a model independent way, whereas in section~\ref{comparison} we will specialise to three popular models. 

\subsection{Kinetic Energy Domination at the Bounce: $\frac{PE_{B}}{\rho_{c}} \ll \frac{1}{2}$}
\label{KE-dom-regime}

\subsubsection{The Superinflation Regime} \label{Superinflation-regime}
\noindent As we leave the bounce we do so with the KE dominating over the PE of $\phi$, i.e. $\frac{\dot{\phi}^{2}}{2} \gg V(\phi)$. Hence the total energy $\rho \approx \frac{\dot{\phi}^{2}}{2}$. We also initially assume that $\ddot{\phi} \gg V'(\phi)$ which is consistent with being dominated by the KE. It follows that equations (\ref{Friedmann}) and (\ref{KG}) in this regime become:

\begin{equation} \label{Friedmann-KEdom}
H^{2} = \frac{4\pi\dot{\phi}^{2} }{3 m^2_{\rm Pl}} \bigg(1-\frac{\dot{\phi}^{2}}{2\rho_{c}}\bigg) \\
\end{equation}
\begin{equation} \label{KG-KEdom}
\ddot{\phi} + 3H\dot{\phi} = 0\\
\end{equation}
Following \cite{Ashtekar-Sloan,Linsefors-Barrau}, these equations can be solved exactly. With initial conditions $(\rho,\dot{\phi},\phi)=(\rho_{c},\pm \sqrt{2\rho_{c}},\phi_{B})$ at $t=0$, we obtain :
\begin{equation} \label{phidot-KEdom}
\dot{\phi}(t) = \pm \frac{\sqrt{2\rho_{c}}}{\sqrt{1 + 24\pi\rho_{c}t^{2}/m^2_{\rm Pl}}} \\
\end{equation}
\begin{equation} \label{phi-KEdom}
\phi(t) =  \phi_{B} \pm \frac{m_{\rm Pl}}{\sqrt{12\pi}}\sinh^{-1}\left(\frac{\sqrt{24\pi\rho_{c}}t}{m_{\rm Pl}}\right) \\
\end{equation}
where the $\pm$ sign results from the choice of the initial field velocity at the bounce (Note that this solution was also obtained earlier for the case of $V(\phi) =0$ \cite{Mielczarek:2008qw}). Equations (\ref{phidot-KEdom}) and (\ref{phi-KEdom}) lead to an early period of superinflation just after the bounce, where, independently of the form of the potential $V(\phi)$ because we are in the KE dominated regime, the scale factor behaves as $a(t) = (1+24\pi \rho_{c} t^{2}/m^2_{\rm Pl})^{1/6}$ with inflation ($\ddot{a}>0$) lasting until $t < (16 \pi \rho_{c})^{-\frac{1}{2}}m_{\rm Pl}$ (corresponding to of order 0.2 e-foldings). During this period the Hubble parameter rapidly increases from zero at the bounce to a maximum value of $(2\pi\rho_{c}/(3m^2_{\rm Pl}))^{1/2}$ in a time $t=(24\pi\rho_{c})^{-1/2}m_{\rm Pl}\approx0.18$ Planck seconds, before it begins to decrease again. This is of course in contrast to potential driven slow roll inflation, where $H \sim$ const during inflation. As stated earlier, because we are in a KE dominated regime, our result is independent of the form of the potential, although we do assume it is regular everywhere.   \\~\\
To a first approximation the KE dominated regime lasts until the kinetic and potential energies are equal. Although of course not strictly true, the fact that the fields evolve quickly to this regime suggests it should be a good approximation, and we confirm that below. Therefore we can substitute for $\phi$ and $\dot{\phi}$ from equations (\ref{phidot-KEdom}-\ref{phi-KEdom}) to yield the time $t_{\rm eq}$, when $\dot{\phi}_{\rm eq}^2 = 2 V(\phi_{\rm eq})$ for a given potential $V(\phi)$. A similar approach to establishing the onset of equilibrium has been adopted in \cite{Zhu New}.  
As an example, for the case of a quadratic inflaton potential ($\frac{1}{2} m^2 \phi^2 $), we see from Table~\ref{Table1}, with $\phi_{B}=1.06 m_{\rm Pl}$, at $t_{\rm eq}$, the value of $(\phi,\dot{\phi})$ predicted by Equations.~(\ref{phidot-KEdom}-\ref{phi-KEdom}) is $(3.19 m_{\rm Pl},3.86\times10^{-6}m^2_{\rm Pl})$ as opposed to the values obtained using the full numerical solutions \cite{Ashtekar-Sloan}, $(3.16 m_{\rm Pl},3.82\times10^{-6}m^2_{\rm Pl})$ with percentage errors of $(0.95,1.05)$.  The total number of e-foldings up to this period is between $N\sim 4-5$ \cite{Ashtekar-Sloan,Bonga-Gupt}. We will see that this is true for all the potentials we use, giving us confidence in the way we discuss the end of superinflation and the onset of the damping regime. As a further justification of the validity of the approximation we have just made, defining the temporal average of a quantity $A(t)$ by $\langle A\rangle=\frac{1}{t_{eq}}\int_{0}^{t_{eq}}A(t)dt$, we find numerically that 
$\frac{\langle V(\phi)\rangle}{\langle{\dot{\phi}^2/2}\rangle}$ is of order $10^{-6}-10^{-7}$ for all the potentials in this regime and moreover $\langle\ddot{\phi}\rangle:\langle 3H\dot{\phi}\rangle:\langle V'\rangle$ is $1:1:10^{-7}$ for quadratic/quartic potentials and $1:1:10^{-9}$ for the Starobinsky potential, thereby confirming the validity of our assumption that $\ddot{\phi} \gg V'(\phi)$ up to $t_{\rm eq}$. The end of this regime gives us $(\phi_{\rm eq},\dot{\phi}_{\rm eq})$ at the equilibrium point which serves as the initial condition for the next phase, the damping phase. 

\subsubsection{The Damping Regime}\label{KEDom-Damping-Phase}
\noindent As we enter the regime where the KE is comparable to the PE, the KE is dropping rapidly compared to the potential contribution and the overall combined energy density is dropping quickly compared to the bounce density $\rho_c$. As a result, we rapidly enter a regime where we are potential dominated, $2 V(\phi) \gg \dot{\phi}^2$, and we can drop the correction term $\frac{\rho}{\rho_c}$ in the modified Friedmann equation (\ref{Friedmann}), which now becomes
\begin{equation} \label{Friedman-damping}
H^{2} = \frac{8\pi}{3m^2_{\rm Pl}}V(\phi) \\
\end{equation}
Given that the field has slowed down, we don't expect it to travel far before it stops and turns around. Therefore we can obtain an accurate estimate of this period by perturbing the field around its equilibirum values $(\phi_{\rm eq},\dot{\phi}_{\rm eq})$ and working to leading order as
\begin{eqnarray}
V(\phi) & = & V(\phi_{\rm eq}) + (\phi-\phi_{\rm eq})V'(\phi_{\rm eq}) + O(\phi-\phi_{\rm eq})^{2} \label{V-damping}\\
V'(\phi) & = & V'(\phi_{\rm eq}) + (\phi-\phi_{\rm eq})V''(\phi_{\rm eq}) + O(\phi-\phi_{\rm eq})^{2} \label{Vprime-damping}
\end{eqnarray}
Substituting Eqns~(\ref{V-damping} - \ref{Vprime-damping}) into Eqns~(\ref{Friedman-damping}) and (\ref{KG}), and expanding to linear order in $\phi(t)$, we obtain after defining 
\begin{equation}\label{psi-definition}
\psi(t) \equiv \phi(t) + (V'(\phi_{\rm eq}) - \phi_{\rm eq} V''(\phi_{\rm eq}))/V''(\phi_{\rm eq}),
\end{equation}
\begin{equation} \label{psi-eqn}
\ddot{\psi} + \alpha \dot{\psi} + \beta \psi  =  0
\end{equation}
where for simplicity we have introduced $\alpha =\sqrt{24\pi V(\phi_{\rm eq})/m^2_{\rm Pl}}$ and  $\beta =V''(\phi_{\rm eq})$. This is the equation for a damped harmonic oscillator and has a general solution: 
\begin{equation} \label{psi-soln}
\psi  =  Ae^{- \lambda_+ t} + Be^{- \lambda_- t}
\end{equation}
where $A$ and $B$ are constants and $\lambda_{\pm}$ are given by 
\begin{equation} \label{eignevalues}
2 \lambda_{\pm} = \alpha \pm \sqrt{\alpha^2 - 4 \beta}.
\end{equation}
We can determine the coefficients $A$ and $B$ from the initial conditions namely the known values of $(\phi_{\rm eq},\dot{\phi}_{\rm eq})$ or equivalently $(\psi_{\rm eq},\dot{\psi}_{\rm eq})$ at $t=0$ (where we have reset the time coordinate so that $t=0$ now corresponds to the onset of the damping regime). For completeness we obtain 
\begin{equation}
A = {\lambda_- \psi_{\rm eq} + \dot{\psi}_{\rm eq} \over \lambda_- - \lambda_+}\;,~~~~~~~~
B = -{(\lambda_+ \psi_{\rm eq} + \dot{\psi}_{\rm eq}) \over \lambda_- - \lambda_+}. \label{Bcoeff} 
\end{equation}
The solution we have obtained in Eqns~(\ref{psi-soln}-\ref{Bcoeff}) is remarkably accurate considering the approximations we have made. Once again, considering the example of the quadratic potential with $\phi_{B}=1.06m_{\rm Pl}$, we can predict the time $t_{\rm turn}$ when the inflaton turns around by differentiating Eqn~(\ref{psi-soln}) and solving $\dot{\psi}=0$. In particular we find that 
\begin{equation}\label{turn-around-condition}
t_{\rm turn} = \frac{1}{\lambda_+-\lambda_-} \ln \left( \frac{\psi_{\rm eq} + \dot{\psi}_{\rm eq}/\lambda_-}{\psi_{\rm eq} + \dot{\psi}_{\rm eq}/\lambda_+}\right)
\end{equation}
For the case where $ \dot{\psi}_{\rm eq} >0$, we obtain $\phi_{\rm turn}=3.32m_{\rm Pl}$ which should be compared with the numerical solution of $\phi_{\rm turn}=3.27m_{\rm Pl}$ (a 1.5\%  error). It follows from Eqn.~(\ref{turn-around-condition}) that if $\dot{\psi}_{\rm eq} <0$, then there will not be a turn around regime. This can easily be seen by recalling that $\lambda_+>\lambda_-$, hence for the case $\dot{\psi}_{\rm eq} <0$, we obtain the inconsistent result that $t_{\rm turn} <0$, hence it can not happen. Returning to the case of $ \dot{\psi}_{\rm eq} >0$, a similar level of accuracy holds for the other potentials. To justify the approximations we have made (dropping the KE term in the Friedmann constraint Eqn~(\ref{Friedman-damping}) yet keeping all the terms in the Klein Gordon equation~(\ref{KG})) we once again look at the averages  
$\frac{\langle{\dot{\phi}^2/2}\rangle}{\langle V(\phi)\rangle}$ and $\langle\ddot{\phi}\rangle:\langle3H\dot{\phi}\rangle:\langle V'(\phi)\rangle$. For all potentials  we investigated the former is around $0.05$ implying that on average the KE density makes up only 5 percent of the total energy density. For the latter case we find $1:1:0.05$ for Starobinsky ($\phi_{B}=3.67m_{\rm Pl}$), $2.72:1.72:1$ for quadratic ($\phi_{B}=1.06m_{\rm Pl}$) and $2:2:1$ for the quartic potential ($\phi_{B}=2.35m_{\rm Pl}$). In other words apart from arguably the Starobinsky case, all the terms are of the same magnitude which justifies retaining all three terms in Eqn.~(\ref{KG}).  \\~\\
As the field slows down, we eventually reach a situation where its acceleration $\ddot{\phi}$ vanishes, and this marks the end of the damping regime. Numerically this is not so easy to determine and so we look for the condition  
$\big|\frac{\ddot{\phi}}{3H\dot{\phi}}\big|=0.01$ to occur the first time after the equilibrium point, using it as a mark of the end of the damping and the onset of the slow roll regime. The value of $\ddot{\phi}$ is simply determined from the continuity Eqn~(\ref{KG}) and the Hubble parameter $H$ from the Friedmann constraint Eqn~(\ref{Friedmann}). At this point the value of $\phi_{SRA},\dot{\phi}_{SRA}$ serves as the initial conditions for the Slow-Roll regime, and of course can be determined analytically from our solutions (\ref{psi-soln}-\ref{Bcoeff}) when $\big|\frac{\ddot{\phi}}{3H\dot{\phi}}\big|=0.01$. We have checked that our results are not sensitive to the precise value $0.01$, it just has to be a lot smaller than unity. 

\subsubsection{The Slow-Roll Regime} \label{slow-roll-regime}
\noindent As we leave the damping regime the only change is that we now have $\ddot{\phi} \approx 0$, hence the Friedmann and continuity equations become
\begin{equation} \label{Friedman-SRA}
H^{2} = \frac{8\pi}{3m^2_{\rm Pl}}V(\phi) \\
\end{equation}
\begin{equation} \label{KG-SRA}
3H\dot{\phi}+V'(\phi) = 0 
\end{equation}
which are in fact the standard slow-roll equations. We can eliminate the Hubble parameter in Eqn~(\ref{KG-SRA}) and solve for $\dot{\phi}$ yielding:
\begin{equation} \label{phidot-SRA}
\dot{\phi} = \frac{-m_{\rm Pl} V'(\phi)}{\sqrt{24\pi V(\phi)}}
\end{equation}
which can be further integrated to determine the evolution of $\phi$ once we specify $V(\phi)$, using the slow-roll point evaluated at the end of the damping regime $(\phi_{\rm SRA}, \dot{\phi}_{\rm SRA}$) as our initial condition. \\~\\
Given that, we then know the value of $\phi_*$, and we can estimate the values of observables at any point in the slow-roll regime such as the pivot-point. It is important to note from all the Tables~\ref{Table1} - \ref{Table4} that if the slow-roll starts before/at the pivot point, then slow-roll inflation compatible with observations will be achieved. It is only in those cases do we apply Eqn.~(\ref{phidot-SRA}) for the pivot-point.  

\subsection{Potential Energy Domination at the Bounce : $\frac{PE_{B}}{\rho_{c}} \gg \frac{1}{2}$} \label{PE-at-bounce}

\subsubsection{The Damping Regime}
\noindent In this case we come out of the bounce straight into a PE dominated regime with $\rho \simeq V(\phi) \leq \rho_{c}$. In such a case the quantum gravity motivated correction to the Friedmann equation is important and Eqn.(\ref{Friedmann}) becomes: 
\begin{equation}
H^{2} = \frac{8\pi}{3m^2_{\rm Pl}}V(\phi)\Bigg[1-\frac{V(\phi)}{\rho_{c}}\Bigg]. \label{Friedmann-PEdom}
\end{equation}
The fact we are in a PE dominated regime implies the velocity of the field is small, $\dot{\phi}^2 \ll 2 V(\phi)$. We can take the same approach as we did with the damping regime in section~\ref{KE-dom-regime}, namely we can Taylor expand the potential to linear order but this time around the point $\phi \simeq \phi_{\rm B}$. This results in a similar set of equations to  (\ref{psi-eqn}-\ref{Bcoeff}) but with $\alpha =\sqrt{\frac{24\pi V(\phi_{\rm B})}{m^2_{\rm Pl}}\left(1-\frac{V(\phi_{\rm B})}{\rho_c}\right)}$,  $\beta =V''(\phi_{\rm B})$ and 
$\psi(t) \equiv \phi(t) + (V'(\phi_{\rm B}) - \phi_{\rm B} V''(\phi_{\rm B}))/V''(\phi_{\rm B})$. 
The coefficients $A$ and $B$ are now obtained from the initial conditions $(\phi_{\rm B},\dot{\phi}_{\rm B})$ or equivalently $(\psi_{\rm B},\dot{\psi}_{\rm B})$ at $t=0$ (where we have reset the time coordinate so that $t=0$ now corresponds to the onset of this damping regime) and are given by 
\begin{equation}
A = {\lambda_- \psi_{\rm B} + \dot{\psi}_{\rm B} \over \lambda_- - \lambda_+}\;,~~~~~~~~B = -{(\lambda_+ \psi_{\rm B} + \dot{\psi}_{\rm B}) \over \lambda_- - \lambda_+}
\end{equation}

\subsubsection{The Slow-Roll Regime}
\noindent Following the damping regime, the field evolves as in section~\ref{KE-dom-regime} in that it slows down eventually stopping and enters the slow roll regime defined by $\big|\frac{\ddot{\phi}}{3H\dot{\phi}}\big|=0.01$. It leads to an equation similar to (\ref{phidot-SRA}) but with the holonomy correction factor in the denominator,
\begin{equation}\label{phidot-SRAholonomy}
\dot{\phi} = \frac{-m_{\rm Pl} V'(\phi)}{\sqrt{24\pi V(\phi)\big(1-\frac{V(\phi)}{\rho_{c}}\big)}}
\end{equation}
the extra term being there because of the fact that the slow-roll starts very close to the bounce and thus there is hardly any change in the value of the potential i.e. the value of potential at the slow-roll point is still very close to the critical energy density. The evolution then follows in a manner similar to that described in section~\ref{slow-roll-regime}.

\section{Comparison of Analytic versus Numerical solutions} \label{comparison}
\noindent Having established the key dynamical regions and associated equations which can take us from the bounce through to the onset of inflation in a model independent way, we now turn our attention to the case of three particularly popular inflation potentials, namely the quadratic and quartic chaotic potentials as well as the Starobinsky potential. Although the chaotic potentials are tightly constrained by the Planck data \cite{Ade:2015xua}, they provide an excellent testbed for our approach.   We will compare our analytic solutions with the exact numerical solutions obtained in \cite{Ashtekar-Sloan,Bonga-Gupt} and show how close they are to each other throughout the evolution (for earlier work on inflationary attractors with $m^2 \phi^2$ potentials in LQC see Ref~\cite{Singh:2006im} and also Ref~\cite{Mielczarek:2009zw}).
  
\subsection{Quadratic : $V(\phi)=\frac{1}{2}m^{2}\phi^{2}$}
\label{quadratic-numerical-analytic}
\noindent The fact that the quadratic potential is symmetric means it has potentially interesting observable features for both positive and negative values of $\phi_{\rm B}$. Applying the procedure discussed in \cite{Bonga-Gupt} we obtain the values of the mass parameter in the potential as well as the  conditions at the pivot point which we remind the reader is 58 efoldings before the end of inflation, to be compatible with observations of the anisotropies in the CMB: $m=1.21\times10^{-6}m_{\rm Pl}$, $\phi_{*}=\pm3.15m_{\rm Pl}$, $\dot{\phi}_{*}=\mp1.97\times10^{-7}m_{\rm Pl}^{2}$ and $\phi_{end}=0.282m_{\rm Pl}$.  \\~\\
In Tables~\ref{Table1}-\ref{Table2} we compare our analytic solutions with the published numerical results. Table~\ref{Table1} corresponds to $\phi_{\rm B} >0$ and Table~\ref{Table2} corresponds to $\phi_{\rm B} <0$, and in both cases we  begin in a KE dominated regime with $\dot{\phi}_B>0$. Considering Table I, since we are KE dominated at the bounce, the inflaton begins its evolution close to the bottom of the potential $\phi_{\rm B} \simeq 0$ with positive velocity $\dot{\phi}_{B}=\sqrt{2(\rho_{c}-V(\phi_{B}))}\approx9.05\times10^{-1} m^2_{\rm Pl}$ and starts to move up its potential. In this regime it undergoes superinflation which ends at $\rho=\rho_{c}/2$ (recall this is when $\dot{H}=0$) where we can use Eqns.~(\ref{phidot-KEdom}-\ref{phi-KEdom}) to determine the values $(\phi,\dot{\phi},t)$ which we denote as 'End of SI' in the tables. As can be seen from Table~\ref{Table1} the values of $\phi(t)$, ${\dot{\phi}}(t)$ and other observables obtained analytically in the different epochs of evolution are close to the ones obtained numerically, although of course, due to the nature of our approximations, differences do emerge. They are most prominent at the end of the Superinflation phase, where we approach the equilibrium point.  \\~\\
For $\phi_{B}=1.06 m_{\rm Pl}$ in Table \ref{Table1} we use the condition $\dot{\phi}_{\rm eq}^2 = 2 V(\phi_{\rm eq})$ to find the time at which KE=PE and then substitute the corresponding $t_{eq}$ into Eqns.~(\ref{phidot-KEdom}) and (\ref{phi-KEdom}) to obtain $(\phi_{eq},\dot{\phi}_{eq})=(3.19 m_{\rm Pl},3.86\times10^{-6}m^2_{\rm Pl})$ which as can be seen agrees very well with the numerical evolution.  \\~\\
Turning now to the damping phase, in Eqn.~(\ref{psi-eqn}), we find $\alpha=\sqrt{24\pi V_{eq}/m^2_{\rm Pl}}=(m/m_{\rm Pl})\sqrt{12\pi}\phi_{eq}=2.4\times10^{-5}m_{\rm Pl}$ and $\beta=V''(\phi_{eq})=1.5\times10^{-12} m^2_{\rm Pl}$  which yield the damping roots in Eqn.~(\ref{eignevalues}), $(\lambda_+,\;\lambda_-)=(-2.36\times10^{-5}m_{\rm Pl},-6.20\times10^{-8}m_{\rm Pl})$. Now we can use Eqn.~(\ref{psi-soln}) and the initial conditions $(\phi_{eq},\dot{\phi}_{eq})$ to obtain the leading contribution to the damped solution 
\begin{equation}\label{damped-soln}
\phi(t) = (-0.17m_{\rm Pl})e^{-2.36\times10^{-5}m_{\rm Pl}t} + (3.36m_{\rm Pl})e^{-6.20\times10^{-8}m_{\rm Pl}t},\\
\end{equation}
where as usual we have reset the time so that $t=0$ corresponds to the initial condition at equality. To find the turnaround point we set $\dot{\phi}(t)=0$ to obtain $t_{Turn}=1.26\times10^{5}m^{-1}_{\rm Pl}$ which in Eqn.~(\ref{damped-soln}) yields $\phi_{{\rm Turn}}=3.32m_{\rm Pl}$ whereas numerically we obtain from the fourth block of Table~\ref{Table1}, $\phi_{{\rm Turn}}=3.27m_{\rm Pl}$, which is in excellent agreement. \\~\\
To find the slow-roll point we demand $\big|\frac{\ddot{\phi}}{3H\dot{\phi}}\big|=0.01$ the first time it happens after turnaround. Using Eqn.~(\ref{damped-soln}) we obtain $(\phi,\dot{\phi})=(3.31m_{\rm Pl},-1.95\times10^{-7}m^2_{\rm Pl})$ compared to the numerical result in Table~\ref{Table1} which is $(3.25m_{\rm Pl},-1.95\times10^{-7}m^2_{\rm Pl})$, again with excellent agreement. One technical point worth mentioning is that although we look for its absolute value, the sign of $\frac{\ddot{\phi}}{3H\dot{\phi}}$ can be either positive or negative. It turns out that for Table~\ref{Table1} we have the positive sign whereas for Table~\ref{Table2} we have the negative sign for the start of the slow roll regime. It is a nice feature of our analytic treatment that we not only match the numerical values for the observables but also their sign, hence the physics of the system. Another physical feature that is described by our analytic treatment is the point where $\ddot{\phi}=0$. In the case where $\phi_{B}>0, \dot{\phi}_{B}>0 $, hence is moving up the potential (as in Table~\ref{Table1}) this condition is indeed satisfied, however in the case where $\phi_{B}>0, \dot{\phi}_{B}<0 $, hence is moving down the potential (as in Table~\ref{Table2}), this condition is never reached. This is why we use the less constraining condition $\big|\frac{\ddot{\phi}}{3H\dot{\phi}}\big|=0.01$ to detemine the onset of slow roll. 
Considering the results in Table~\ref{Table1}, we see that for $\phi_{B}=1.06m_{\rm Pl}$ we obtain $\phi_{SRA}=3.31m_{\rm Pl}>\phi_{*}=3.15m_{\rm Pl}$, thus the regime where the slow-roll is compatible with  observations is satisfied and we can use Eqn.~(\ref{phidot-SRA}) to determine the key observables in the slow-roll regime. For example we find that the inflaton velocity at the pivot point is $\dot{\phi}=-(mm_{\rm Pl})/\sqrt{12\pi}=-1.97\times10^{-7}m^2_{\rm Pl}$. The situation for the results in Table~\ref{Table2} is similar except that now the inflaton simply starts rolling down the hill from the beginning and thus after the equilibrium point, it never turns around, instead approaching slow-roll directly. Here again the analytics agree with the published numerics to within a percent or so. \\~\\
Once we have calculated $\phi$ and $\dot{\phi}$ we can determine the Hubble parameter $H$ and $\ddot{\phi}$ in the relevant regimes by using the equations of motion appropriate to those regimes, In particular we can obtain $H$ from Eqn.~(\ref{Friedmann}) and $\ddot{\phi}$ from Eqn.~(\ref{KG}) by substituting in the appropriate solutions for $\phi$ and $\dot{\phi}$. 

\begin {table}[H] 
\begin{center}
	\begin{tabular}{ |c||c|c|c||c|c|c| } 
		\hline
		Event & $\phi_{A}/m_{\rm Pl}$ & $\dot{\phi_{A}}/m^2_{\rm Pl}$ & $t_{A} m_{\rm Pl}$ & $\phi_{N}/m_{\rm Pl}$ & $\dot{\phi_{N}}/m^2_{\rm Pl}$ & $t_{N}m_{\rm Pl}$ \\
		\hline 
		i. Bounce & $0.800$ & $9.05\times10^{-1}$ & $0$ & $0.800$ & $9.05\times10^{-1}$ & $0$ \\ 
		End of SI & $0.943$ & $6.40\times10^{-1}$ & $1.80\times10^{-1}$ & $0.943$ & $6.40\times10^{-1}$ & $1.80\times10^{-1}$ \\ 
		KE = PE & $2.94$ & $3.56\times10^{-6}$ & $4.58\times10^{4}$ & $2.91$ & $3.52\times10^{-6}$ & $4.04\times10^{4}$ \\
		Turn & $3.08$ & $0$ & $1.79\times10^{5}$ & $3.03$ & $0$ & $1.65\times10^{5}$ \\
		Slow-Roll & $3.06$ & $-1.95\times10^{-7}$ & $3.15\times10^{5}$ & $2.99$ & $-1.95\times10^{-7}$ & $3.71\times10^{5}$ \\
		$*$ & TFE & TFE & TFE & TFE & TFE & TFE \\
		\hline
		ii. Bounce & $0.898$ & $9.05\times10^{-1}$ & $0$ & $0.898$ & $9.05\times10^{-1}$ & $0$ \\ 
		End of SI & $1.04$ & $6.40\times10^{-1}$ & $1.80\times10^{-1}$ & $1.04$ & $6.40\times10^{-1}$ & $1.80\times10^{-1}$ \\ 
		KE = PE & $3.03$ & $3.67\times10^{-6}$ & $4.44\times10^{4}$ & $3.00$ & $3.63\times10^{-6}$ & $3.91\times10^{4}$ \\
		Turn & $3.17$ & $0$ & $1.75\times10^{5}$ & $3.12$ & $0$ & $1.62\times10^{5}$ \\
		Slow-Roll & $3.15$ & $-1.95\times10^{-7}$ & $3.08\times10^{5}$ & $3.09$ & $-1.95\times10^{-7}$ & $3.61\times10^{5}$  \\
		$*$ &  $3.15$ & $-1.97\times10^{-7}$ & $3.11\times10^{5}$ & TFE & TFE & TFE \\
		\hline
		iii. Bounce & $0.961$ & $9.05\times10^{-1}$ & $0$ & $0.961$ & $9.05\times10^{-1}$ & $0$ \\ 
		End of SI & $1.10$ & $6.40\times10^{-1}$ & $1.80\times10^{-1}$ & $1.10$ & $6.40\times10^{-1}$ & $1.80\times10^{-1}$ \\ 
		KE = PE & $3.09$ & $3.74\times10^{-6}$ & $4.35\times10^{4}$ & $3.06$ & $3.70\times10^{-6}$ & $3.84\times10^{4}$ \\
		Turn & $3.23$ & $0$ & $1.72\times10^{5}$ & $3.18$ & $0$ & $1.59\times10^{5}$ \\
		Slow-Roll & $3.21$ & $-1.95\times10^{-7}$ & $3.03\times10^{5}$ & $3.15$ & $-1.95\times10^{-7}$ & $3.55\times10^{5}$ \\
		$*$ & $3.15$ & $-1.97\times10^{-7}$ & $6.13\times10^{5}$ & $3.15$ & $-1.95\times10^{-7}$ & $3.56\times10^{5}$  \\
		\hline
		iv. Bounce & $1.06$ & $9.05\times10^{-1}$ & $0$ & $1.06$ & $9.05\times10^{-1}$ & $0$ \\ 
		End of SI & $1.20$ & $6.40\times10^{-1}$ & $1.80\times10^{-1}$ & $1.20$ & $6.40\times10^{-1}$ & $1.80\times10^{-1}$ \\ 
		KE = PE & $3.19$ & $3.86\times10^{-6}$ & $4.22\times10^{4}$ & $3.16$ & $3.82\times10^{-6}$ & $3.73\times10^{4}$ \\
		Turn & $3.32$ & $0$ & $1.68\times10^{5}$ & $3.27$ & $0$ & $1.56\times10^{5}$ \\
		Slow-Roll & $3.31$ & $-1.95\times10^{-7}$ & $2.96\times10^{5}$ & $3.25$ & $-1.95\times10^{-7}$ & $3.46\times10^{5}$ \\
		$*$ & $3.15$ & $-1.97\times10^{-7}$ & $1.11\times10^{6}$ & $3.15$ & $-1.97\times10^{-7}$ & $8.33\times10^{5}$ \\
		\hline 
	\end{tabular} 
	\caption{Comparison of Analytic (A) and Numerical (N) \cite{Ashtekar-Sloan} results for the evolution of observables for KE dominated cases with $\phi_{B}>0$ for a quadratic potential, $V(\phi)=\frac{1}{2}m^{2}\phi^{2}$ where $m=1.21\times10^{-6} m_{\rm Pl}$. Each block (labelled i -- iv) represents an initial condition at the bounce and each row within a block represents an event in ascending order: Bounce, End of Superinflation, Equilibrium, Turnaround, Slow-roll and Pivot point. An observationally consistent period of slow-roll happens when $\phi_{B} \geq 0.898m_{\rm Pl}$ for the analytic and $\phi_{B} \geq 0.961m_{\rm Pl}$ for the numerical simulations. TFE means too few e-foldings}\label{Table1}
\end{center} 
\end{table}

\begin {table}[H]
\begin{center}
	\begin{tabular}{ |c||c|c|c||c|c|c| } 
		\hline
		Event & $\phi_{A}/m_{\rm Pl}$ & $\dot{\phi_{A}}/m^2_{\rm Pl}$ & $t_{A} m_{\rm Pl}$ & $\phi_{N}/m_{\rm Pl}$ & $\dot{\phi_{N}}/m^2_{\rm Pl}$ & $t_{N}m_{\rm Pl}$ \\
		\hline 
		i. Bounce & $-5.40$ & $9.05\times10^{-1}$ & $0$ & $-5.40$ & $9.05\times10^{-1}$ & $0$ \\ 
		End of SI & $-5.26$ & $6.40\times10^{-1}$ & $1.80\times10^{-1}$ & $-5.26$ & $6.40\times10^{-1}$ & $1.80\times10^{-1}$ \\ 
		KE = PE & $-3.28$ & $3.97\times10^{-6}$ & $4.11\times10^{4}$  & $-3.31$ & $4.00\times10^{-6}$ & $3.62\times10^{4}$ \\		
		Slow-Roll & $-3.08$ & $1.99\times10^{-7}$ & $2.74\times10^{5}$ & $-3.11$ & $1.99\times10^{-7}$ & $3.47\times10^{5}$ \\
		$*$ & TFE & TFE & TFE & TFE & TFE & TFE  \\
		\hline 
		ii. Bounce & $-5.44$ & $9.05\times10^{-1}$ & $0$ & $-5.44$ & $9.05\times10^{-1}$ & $0$ \\ 
		End of SI & $-5.30$ & $6.40\times10^{-1}$ & $1.80\times10^{-1}$ & $-5.30$ & $6.40\times10^{-1}$ & $1.80\times10^{-1}$ \\ 
		KE = PE & $-3.32$ & $4.02\times10^{-6}$ & $4.05\times10^{4}$ & $-3.35$ & $4.05\times10^{-6}$ & $3.57\times10^{4}$ \\		
		Slow-Roll & $-3.12$ & $1.99\times10^{-7}$ & $2.71\times10^{5}$ & $-3.15$ & $1.99\times10^{-7}$ & $3.43\times10^{5}$ \\
		$*$ & TFE & TFE & TFE & $-3.15$ & $1.98\times10^{-7}$ & $3.57\times10^{5}$ \\
		\hline 
		iii. Bounce & $-5.47$ & $9.05\times10^{-1}$ & $0$ & $-5.47$ & $9.05\times10^{-1}$ & $0$ \\ 
		End of SI & $-5.33$ & $6.40\times10^{-1}$ & $1.80\times10^{-1}$ & $-5.33$ & $6.40\times10^{-1}$ & $1.80\times10^{-1}$ \\ 
		KE = PE & $-3.35$ & $4.05\times10^{-6}$ & $4.02\times10^{4}$ & $-3.38$ & $4.09\times10^{-6}$ & $3.54\times10^{4}$ \\		
		Slow-Roll & $-3.15$ & $1.99\times10^{-7}$ & $2.70\times10^{5}$ & $-3.18$ & $1.99\times10^{-7}$ & $3.40\times10^{5}$ \\
		$*$ & $-3.15$ & $1.97\times10^{-7}$ & $2.85\times10^{5}$ & $-3.15$ & $1.97\times10^{-7}$ & $5.17\times10^{6}$ \\
		\hline
		iv. Bounce & $-5.50$ & $9.05\times10^{-1}$ & $0$ & $-5.50$ & $9.05\times10^{-1}$ & $0$ \\ 
		End of SI & $-5.36$ & $6.40\times10^{-1}$ & $1.80\times10^{-1}$ & $-5.36$ & $6.40\times10^{-1}$ & $1.80\times10^{-1}$ \\ 
		KE = PE & $-3.38$ & $4.09\times10^{-6}$ & $3.98\times10^{4}$ & $-3.41$ & $4.13\times10^{-6}$ & $3.51\times10^{4}$ \\		
		Slow-Roll & $-3.18$ & $1.99\times10^{-7}$ & $2.68\times10^{5}$ & $-3.22$ & $1.99\times10^{-7}$ & $3.38\times10^{5}$ \\
		$*$ & $-3.15$ & $1.97\times10^{-7}$ & $4.20\times10^{5}$ & $-3.15$ & $1.97\times10^{-7}$ & $6.76\times10^{5}$ \\
		\hline
	\end{tabular}
	\caption{Comparison of Analytic (A) and Numerical (N) \cite{Ashtekar-Sloan} results for the evolution of observables for KE dominated cases with $\phi_{B}<0$ for a quadratic potential, $V(\phi)=\frac{1}{2}m^{2}\phi^{2}$ where $m=1.21\times10^{-6}m_{\rm Pl}$. Each block (labelled i -- iv) represents an initial condition at the bounce and each row within a block represents an event in ascending order: Bounce, End of Superinflation, Equilibrium, Slow-roll and Pivot point. An observationally consistent period of slow-roll happens when $\phi_{B}\leq -5.47m_{\rm Pl}$ for the analytic and $\phi_{B}\leq -5.44m_{\rm Pl}$ for the numerical simulations. TFE means too few e-foldings} \label{Table2}
\end{center}
\end{table}
\noindent It is worth noting that there is a difference between the analytic and numerical results when we ask what is the minimum value $\phi$ can have at the bounce that will lead to an observationally consistent period of slow roll inflation. We see from Table~\ref{Table1} ($\phi_B >0$) that analytically we obtain successful evolution for $\phi_B \geq 0.898m_{\rm Pl}$, whereas numerically it is $\phi_B \geq 0.961m_{\rm Pl}$. From Table~\ref{Table2} ($\phi_B <0$), the corresponding numbers are $\phi_B \leq -5.47m_{\rm Pl}$ and $\phi_B \leq -5.44m_{\rm Pl}$ respectively. The difference in these values is a reflection on the approximations being used to obtain the analytic solutions and is to be expected. In actual fact the values are very close to one another (within 6\% and 0.5\% for the two cases), and the real purpose of the analytic approach is to demonstrate how the different regimes can be described and connected to one another for a given initial bounce. \\~\\
In these tables we have not shown any PE dominated cases because in those situations the behaviour is very much as in slow roll inflation. In particular, for the case where $\phi_{B}>0, \dot{\phi}_{B}>0$ the field hardly evolves from the bounce point to turnaround to the slow-roll point. As an example, if at the bounce $\phi_{B}=6\times10^5 m_{\rm Pl}$ then at the turnaround point we will have $\phi_{Turn}=\phi_{B} + 1.00 m_{\rm Pl}$ which is a change of $1.7\times10^{-4}$ percent. Thus the inflaton hardly moves and this agrees with the numerics as is discussed in \cite{Ashtekar-Sloan}. The same is true for the field if initially $\phi_{B}<0, \dot{\phi}_{B}>0$ or $\phi_{B}>0, \dot{\phi}_{B}<0$. The field immediately enters a period of slow roll inflation and standard slow roll inflation continues. 

\subsection{Starobinsky : $V(\phi)=\frac{3M^{2}m^2_{\rm Pl}}{32\pi}\bigg(1-e^{-\sqrt{\frac{16\pi}{3m^2_{\rm Pl}}}\phi}\bigg)^{2}$} \label{Starobinsky-section}
\noindent Unlike the quadratic potential, the Starobinsky potential is not even: close to the origin it behaves quadratically, for large positive values of $\phi$ it asymptotes to a constant, and for large negative values it grows exponentially with $\phi$. Applying the procedure presented in \cite{Bonga-Gupt} we obtain identical values for both the mass parameter in the potential and the necessary conditions at the pivot point to be compatible with observations as: $M=2.51\times10^{-6}m_{\rm Pl}$, $\phi_{*}=1.08m_{\rm Pl}$, $\dot{\phi}_{*}=-4.80\times10^{-9}m_{\rm Pl}^{2}$ and $\phi_{end}=0.187m_{\rm Pl}$ .  \\~\\
Our numerical and analytical results are summarised in Table~\ref{Table3} ($\dot{\phi}_{B}>0$) and Table~\ref{Table4} ($\dot{\phi}_{B}<0$). Blocks (ii-iv) in both tables are for the case when we are KE dominated at the bounce and block (i) in both tables are for the PE dominated case. The KE dominated regimes in both tables are very similar to that discussed in section~\ref{quadratic-numerical-analytic}. In particular (considering Table~\ref{Table3}) the inflaton starts at the bounce, undergoes superinflation and reaches the equilibrium point. For example if $\phi_{B}=-1.37m_{\rm Pl}$ with $\dot{\phi}_{B}>0$, then using Eqns.~(\ref{phidot-KEdom}) and (\ref{phi-KEdom}) we obtain $(\phi_{eq},\dot{\phi}_{eq},t_{eq})=(1.06m_{\rm Pl},6.09\times10^{-7}m^2_{\rm Pl},2.67\times10^{5}m^{-1}_{\rm Pl})$ which agrees with the numerical result  $(1.03m_{\rm Pl},6.04\times10^{-7}m^2_{\rm Pl},2.37\times10^{5}m^{-1}_{\rm Pl})$. Using these as the initial conditions for the damping phase we again calculate the damping roots $\lambda_+$ and $\lambda_-$ in Eqn.~(\ref{eignevalues}), and thus obtain the damped solution for $\phi(t)$ in Eqn.~(\ref{psi-soln}). This predicts the turnaround to happen at $\phi_{Turn}=1.22m_{\rm Pl}$, remarkably close again to the numerical prediction of $\phi_{Turn}=1.17m_{\rm Pl}$. Similarly for the slow-roll point we use the condition $\big|\frac{\ddot{\phi}}{3H\dot{\phi}}\big|=0.01$ to find $t_{SRA}$ which when substituted into Eqn.~(\ref{phidot-SRA}) determines the initial values of the field at the start of slow-roll to be $(\phi,\dot{\phi})=(1.08m_{\rm Pl},-4.81\times10^{-9}m^2_{\rm Pl})$ compared to the numerical values of $(1.16m_{\rm Pl},-3.48\times10^{-9}m^2_{\rm Pl})$. We may use Eqn.~(\ref{phidot-SRA}) to predict the observables at the pivot point if the slow-roll starts by the time we reach it (i.e. if $\phi_{SRA} \geq \phi_*$). One might ask can we obtain the usual slow roll related inflation associated with Starobinsky inflation, when we include the constraints arising from the initial bounce? The answer is yes, as can be seen from Table~\ref{Table4}, but there are some subtleties. The numerical solutions show that slow roll inflation can occur for $\phi_B > 3.63m_{\rm Pl}$, and the analytic approximations have it at $\phi_B > 3.68m_{\rm Pl}$, close but not quite the same. There is a crucial difference between ordinary slow roll inflation and inflation arising in LQC here. In the former case, we can begin on the plateau of the potential and be in the slow roll regime. Inflation can begin almost immediately in such a situation for $\phi > \phi_* = 1.08m_{\rm Pl}$ with 58 e-foldings of inflation. However in the case of LQC, the bounce takes place into a KE dominated regime. The field begins to evolve down its potential and needs to lose this large KE before it can enter a slow roll regime. This implies that $\phi_B$ must be large enough in order to enable the field to lose that KE before entering slow roll. We find that this implies $\phi_B > 3.68m_{\rm Pl}$. One final comment related to the results here, is that the discrepancy between the analytic and numerical evaluations in the velocities at the onset of slow-roll are a reflection of the approximations that have been made.  \\~\\
Now let us discuss the PE dominated cases which correspond to block (i) in Tables~\ref{Table3} and {\ref{Table4}. As can be seen we have $\phi_{B}=-3.47m_{\rm Pl}$ and $\phi_{B}=-3.39m_{\rm Pl}$ respectively. In principle we could apply the techniques discussed in section~\ref{PE-at-bounce} but these are not very helpful in determining whether or not they lead to a period of slow-roll. Instead, we introduce a new approach to determine the solutions in this regime. Considering initially the case of $\dot{\phi}_{B}>0$ (i.e. Table~\ref{Table3}), we have $\phi_{B}=-3.47m_{\rm Pl}$. Using the  superinflation-assisted damping solution presented after Eqn.~(\ref{Friedmann-PEdom}), we find at the end of superinflation $(\phi,\dot{\phi})=(-3.28m_{\rm Pl},0.484m^2_{\rm Pl})$, while the numerical results give $(-3.25m_{\rm Pl},0.520m^2_{\rm Pl})$. By the time this stage is reached, we are entering a regime where the total energy density becomes negligible with respect to the critical density $\rho \ll \rho_{c}$. This then allows us to make use of the scaling behaviour of $\phi$ in this regime, and obtain accurate solutions for its evolution. To see this we note that when $\phi = -3.28m_{\rm Pl}$, the potential can be accurately approximated by $V(\phi)=\frac{3M^{2}m^2_{\rm Pl}}{32\pi}\big(1-e^{-\sqrt{\frac{16\pi}{3m^2_{\rm Pl}}}\phi}\big)^{2}\approx V_{0}e^{-2b\phi}$ where $V_{0}=\frac{3M^{2}m^2_{\rm Pl}}{32\pi}$ and $b=\sqrt{\frac{16\pi}{3m^2_{\rm Pl}}}$.  In this regime, since $\rho \ll \rho_c$, Eqn.~(\ref{Friedmann}) becomes 
\begin{equation} \label{H-evoln-eqn-ansatz}
H=\sqrt{\frac{8\pi V_{0}}{3m^2_{\rm Pl}}}e^{-b\phi}\sqrt{1+\frac{\dot{\phi}^2}{2V(\phi)}} 
\end{equation}
and Eqn~(\ref{KG}) yields 
\begin{equation} \label{phi-evoln-eqn-ansatz}
\ddot{\phi}+3\sqrt{\frac{8\pi V_{0}}{3m^2_{\rm Pl}}}\sqrt{1+\frac{\dot{\phi}^{2}}{2V(\phi)}}e^{-b\phi}\dot{\phi}-2bV_{0}e^{-2b\phi} = 0
\end{equation}
The crucial observation is that in this regime $\phi$ satisfies the following equation 
\begin{equation} \label{scaling-eqn}
\dot{\phi} = \gamma e^{-b \phi}
\end{equation}
where $\gamma = {\rm const}$. This can be seen by substituting Eqn.~(\ref{scaling-eqn}) into Eqn.~(\ref{phi-evoln-eqn-ansatz}) with $V(\phi) = V_{0}e^{-2b\phi}$ yielding an algebraic equation for $\gamma$
\begin{equation} \label{gamma-soln}
-\gamma^{2}b+\sqrt{\frac{24\pi V_{0}}{m^2_{\rm Pl}}}\sqrt{1+\frac{\gamma^{2}}{2V_{0}}}\gamma-2bV_{0} = 0
\end{equation}
whose solution is $\gamma=5.48\times10^{-7} m^2_{\rm Pl}$ given that $M=2.51\times10^{-6}m_{\rm Pl}$ for this case. Given the initial condition that $\phi(t=0)=\phi_i=-3.28m_{\rm Pl}$ (once again we reset the time coordinate), the solution to Eqn.~(\ref{scaling-eqn}) is 
\begin{equation}\label{phi-soln-early}
\phi(t) = \frac{1}{b} \ln (e^{b\phi_i} + \gamma b t)
\end{equation}
It is worth pointing out that this PE dominated regime is in itself not an inflationary one. This can be seen quickly by recalling that inflation means $\ddot{a}>0$, or equivalently $\dot{H} + H^2 >0$. It follows from Eqns.~(\ref{H-evoln-eqn-ansatz}), (\ref{scaling-eqn}) and (\ref{gamma-soln}) that $\dot{H} + H^2 <0$ in this scaling regime. Using Eqn.~(\ref{scaling-eqn}) we can determine the value of $\phi$ when we are at equilibrium with $\dot{\phi}^{2}/2=V(\phi)$. Care needs to be taken at this point, as we need to use the full potential $V(\phi) = V_0 (1-e^{-b \phi})^2$ to obtain the estimate, 
\begin{equation}\label{phieq}
\phi_{eq}(t) = \frac{1}{b} \ln \left( 1 -  \sqrt{\frac{\gamma^2}{2V_0}}\right)
\end{equation}
In particular we obtain the equilibrium point $(\phi_{eq},\dot{\phi}_{eq},t_{eq})=(-0.55m_{\rm Pl},5.19\times10^{-6}m^2_{\rm Pl},4.70\times10^{4}m^{-1}_{\rm Pl})$ which is remarkably close to the exact numerical values $(-0.55m_{\rm Pl},5.14\times10^{-6}m^2_{\rm Pl},4.70\times10^{4}m^{-1}_{\rm Pl})$. We can estimate when the solution begins to break down, by considering when the approximation for $V(\phi)$ begins to break down. This is when $e^{-b \phi_{\rm br}} \simeq 2$ or $\phi_{\rm br} \simeq - 0.169m_{\rm Pl}$. It turns out that the breakdown of this scaling regime corresponds to the onset of KE domination as the field begins to probe the minimum of its potential. We can solve the system in that regime too. Once again recalling that we are in the regime $\rho \ll \rho_c$, from Eqn.~(\ref{Friedmann})  we have 
\begin{equation} \label{H-evoln-eqn-KE-dom}
H=\sqrt{\frac{4\pi}{3m^2_{\rm Pl}}}\dot{\phi} 
\end{equation}
Similarly Eqn~(\ref{KG}) becomes 
\begin{equation} \label{phi-evoln-eqn-KE-dom}
\ddot{\phi}+3H \dot{\phi}= 0
\end{equation}
with solution 
\begin{equation}\label{phi-KE-dom}
\phi(t) = \phi(0) + \frac{m_{\rm Pl}}{\sqrt{12 \pi}} \ln \left( 1 + \frac{2\sqrt{12 \pi} \gamma t}{m_{\rm Pl}}\right)
\end{equation}
We have used as our initial conditions the values just obtained at the break down point, namely $\phi(0)= -\frac{\ln 2}{b}$ and $\dot{\phi}(0)= 2 \gamma$ (resetting t=0 again in this regime). This solution should be valid as $\phi$ passes through its origin and we use it to estimate when the KE and PE are once again equal on the other side of the potential. Equating the two (recall we use the full $V(\phi)$) we find that 
at this equilibrium point, $(\phi,\dot{\phi})=(0.101m_{\rm Pl},2.08\times10^{-7}m^2_{\rm Pl})$ as opposed to the numerical result $(0.093m_{\rm Pl},1.93\times10^{-7}m^2_{\rm Pl})$ a result accurate to within 8\%. \\~\\
This now supplies us with the initial conditions required as we move into the damped regime which we have discussed in section~\ref{KEDom-Damping-Phase}. We are interested in the value of $\phi$ when it turns around, i.e. $\dot{\phi}=0$, as we need to compare it to the pivot point $\phi_* = 1.08m_{\rm Pl}$. Using Eqns.~(\ref{psi-definition})-(\ref{Bcoeff}), with $(\phi_{\rm eq},\dot{\phi}_{\rm eq})=(0.101m_{\rm Pl},2.08\times10^{-7}m^2_{\rm Pl})$ we find that the inflaton field turns around at $\phi=0.141m_{\rm Pl}$ (again remarkably close to the numerical result of $\phi=0.127m_{\rm Pl}$) i.e. well before it could reach $\phi_{*}=1.08m_{\rm Pl}$. Thus in this case slow-roll will never happen, a result which is consistent with the numerical simulations. \\~\\
We now consider the case of $\dot{\phi}_{B}<0$ (i.e.Table~\ref{Table4}). In this case we have $\phi_{B}=-3.39m_{\rm Pl}$ and we start by going uphill in the potential. The maximum height allowed (by the condition $\rho \leq \rho_c$) is $\phi=-3.47m_{\rm Pl}$ which is found in this PE dominated regime by equating the Starobinsky potential to the critical energy density. We want to estimate the height reached and to do this we make use of the fact that at the bounce $H=0$, and just after the bounce $\dot{\phi} \sim 0$, as it travels up the potential slowing eventually reaching its turn around value, hence for the first few Planck seconds after the bounce the friction term in Eqn.~(\ref{KG}), namely $3H\dot{\phi} \sim 0$. This implies that over this short period, the energy is approximately conserved such that $\frac{1}{2}\dot{\phi}^2 + V(\phi) = \rho_c$. Moreover, given that at the bounce  
 $\phi_{B}=-3.39m_{\rm Pl}$, once again we can say that the potential is well approximated by $V(\phi)=V_0\big(1-e^{-b\phi}\big)^{2}\approx V_{0}e^{-2b\phi}$. It then follows that 
\begin{equation} \label{phidot-energy-con}
\dot{\phi}=-\sqrt{2\rho_{c}-2V_{0}e^{-2b\phi}}
\end{equation}
which can be integrated to yield
\begin{equation}\label{phi-soln-energy-con}
\phi(t) = \frac{1}{b} \ln\left(\sqrt{\frac{V_0}{\rho_c}} \cosh \left[\cosh^{-1}\left(\sqrt{\frac{\rho_c}{V_0}}e^{b \phi_i}\right)-\sqrt{2 \rho_c}  b~t\right]\right)
\end{equation}
where $\phi_i \equiv \phi(t=0)$ is the initial value for the field which in this case is the value at the bounce, namely $\phi_i = \phi_B=-3.39m_{\rm Pl}$. From Eqn.~(\ref{phidot-energy-con}) we see that turnaround ($\dot{\phi}=0$) occurs when $\phi_{Turn} = \frac{1}{2b} \ln (\frac{V_0}{\rho_c})=-3.47m_{\rm Pl}$, which compares favourably with the numerical result $\phi=-3.46m_{\rm Pl}$, showing that the approximation of no friction works well as the field evolves up the potential. Once turn around is reached, $\phi$ begins to evolve down the potential again, and soon we are in a similar position to the situation just described where we entered the scaling regime described in Eqn.~(\ref{scaling-eqn}). Given that this applies to $\phi =-3.39m_{\rm Pl}$ (but now $\dot{\phi}>0$) we see immediately that we are in the same scenario as before and we are unable to reach the required pivot point value $\phi_*$ before the evolution turns around again, a result consistent with the numerical simulations of \cite{Bonga-Gupt} as seen in block (i) of Table~\ref{Table4}. \\~\\
We conclude with the result that in general for PE dominated cases of the Starobinsky potential we are unable to obtain standard slow-roll inflation (a similar conclusion has been reached for the same potential in a class of modified LQC models in Ref~\cite{Li:2018fco}), but recall that we can obtain the standard slow roll behaviour as long as we start in the KE dominated regime with larger values of $\phi_B$, as can be seen in Table~\ref{Table4}.  

\begin {table}[H]
\begin{center}
	\begin{tabular}{ |c||c|c|c||c|c|c| } 
		\hline
		Event & $\phi_{A}/m_{\rm Pl}$ & $\dot{\phi_{A}}/m^2_{\rm Pl}$ & $t_{A} m_{\rm Pl}$ & $\phi_{N}/m_{\rm Pl}$ & $\dot{\phi_{N}}/m^2_{\rm Pl}$ & $t_{N}m_{\rm Pl}$ \\
		\hline 
		i. Bounce & $-3.47$ & $5.07\times10^{-2}$ & $0$ & $-3.47$ & $5.07\times10^{-2}$ & $0$ \\ 
		End of SI & $-3.28$ & $4.84\times10^{-1}$ & $0.41$ & $-3.25$ & $5.20\times10^{-1}$ & $0.46$ \\ 
		KE = PE & $-0.55$ & $5.19\times10^{-6}$ & $4.70\times10^{4}$ & $-0.55$ & $5.14\times10^{-6}$ & $4.70\times10^{4}$ \\
		End of Scaling & $-0.169$ & $1.10\times10^{-6}$ & $2.22\times10^{5}$ & $-0.169$ & $1.062\times10^{-6}$ & $2.25\times10^{5}$ \\
		KE=PE & $0.101$ & $2.08\times10^{-7}$ & $8.55\times10^{5}$ & $0.093$ & $1.93\times10^{-7}$ & $7.64\times10^{5}$ \\
		Turn & $0.141$ & $0$ & $1.27\times10^{6}$ & $0.127$ & $0$ & $1.15\times10^6$ \\
		$*$ & TFE & TFE & TFE & TFE & TFE & TFE \\
		\hline
		ii. Bounce & $-1.45$ & $9.05\times10^{-1}$ & $0$ & $-1.45$ & $9.05\times10^{-1}$ & $0$ \\ 
		End of SI & $-1.31$ & $6.40\times10^{-1}$ & $1.80\times10^{-1}$ & $-1.31$ & $6.40\times10^{-1}$ & $1.80\times10^{-1}$ \\ 
		KE = PE & $0.98$ & $6.08\times10^{-7}$ & $2.68\times10^{5}$ & $0.95$ & $6.01\times10^{-7}$ & $2.38\times10^{5}$ \\
		Turn & $1.14$ & $0$ & $1.73\times10^{6}$ & $1.08$ & $0$ & $1.49\times10^{6}$ \\
		Slow-Roll & $1.00$ & $-6.80\times10^{-9}$ & $4.30\times10^{6}$ & $1.08$ & $-4.88\times10^{-9}$ & $2.91\times10^{6}$  \\
		$*$ & TFE & TFE & TFE & $1.08$ & $-4.85\times10^{-9}$ & $2.76\times10^{6}$ \\
		\hline
		iii. Bounce & $-1.41$ & $9.05\times10^{-1}$ & $0$ & $-1.41$ & $9.05\times10^{-1}$ & $0$ \\ 
		End of SI & $-1.27$ & $6.40\times10^{-1}$ & $1.80\times10^{-1}$ & $-1.27$ & $6.40\times10^{-1}$ & $1.80\times10^{-1}$ \\ 
		KE = PE & $1.02$ & $6.08\times10^{-7}$ & $2.68\times10^{5}$ & $0.99$ & $6.02\times10^{-7}$ & $2.38\times10^{5}$ \\
		Turn & $1.18$ & $0$ & $1.77\times10^{6}$ & $1.12$ & $0$ & $1.53\times10^{6}$ \\
		Slow-Roll & $1.04$ & $-5.71\times10^{-9}$ & $3.88\times10^{7}$ & $1.12$ & $-4.12\times10^{-9}$ & $2.92\times10^{6}$ \\
		$*$ & TFE & TFE & TFE & $1.08$ & $-4.89\times10^{-9}$ & $1.20\times10^{7}$ \\
		\hline
		iv. Bounce & $-1.37$ & $9.05\times10^{-1}$ & $0$ & $-1.37$ & $9.05\times10^{-1}$ & $0$ \\ 
		End of SI & $-1.23$ & $6.40\times10^{-1}$ & $1.80\times10^{-1}$ & $-1.23$ & $6.40\times10^{-1}$ & $1.80\times10^{-1}$ \\ 
		KE = PE & $1.06$ & $6.09\times10^{-7}$ & $2.67\times10^{5}$ & $1.03$ & $6.04\times10^{-7}$ & $2.37\times10^{5}$ \\
		Turn & $1.22$ & $0$ & $1.81\times10^{6}$ & $1.17$ & $0$ & $1.57\times10^{6}$ \\
		Slow-Roll & $1.08$ & $-4.81\times10^{-9}$ & $4.52\times10^{7}$ & $1.16$ & $-3.48\times10^{-9}$ & $2.93\times10^{7}$ \\
		$*$ & $1.08$ & $-4.92\times10^{-9}$ & $4.58\times10^{7}$ & $1.08$ & $-4.89\times10^{-9}$ & $2.28\times10^{7}$ \\
		\hline
	\end{tabular}
	\caption{Comparison of Analytic (A) and Numerical (N) \cite{Bonga-Gupt} results for the evolution of observables with $\dot{\phi}_{B}>0$ for the Starobinsky potential. The last three blocks (ii-iv) are KE dominated cases wherein each row represents an event in ascending order: Bounce, End of Superinflation, Equilibrium, Turnaround, Slow-roll and Pivot point. The first block (i) is for the PE dominated case and contains more rows representing more events. In particular `End of Scaling' refers to when the solution Eqn.~(\ref{scaling-eqn}) breaks down. It does not lead to the desired slow-roll, which instead only occurs in the KE dominated regime with $\phi_{B}\geq -1.37m_{\rm Pl}$ for the analytic and $\phi_{B} \geq -1.45m_{\rm Pl}$ for the numerical cases. TFE means too few e-foldings} \label{Table3}
\end{center}
\end{table}

\begin {table}[H]
\begin{center}
	\begin{tabular}{ |c||c|c|c||c|c|c| } 
		\hline
		Event & $\phi_{A}/m_{\rm Pl}$ & $\dot{\phi_{A}}/m^2_{\rm Pl}$ & $t_{A} m_{\rm Pl}$ & $\phi_{N}/m_{\rm Pl}$ & $\dot{\phi_{N}}/m^2_{\rm Pl}$ & $t_{N}m_{\rm Pl}$ \\
		\hline 
		i. Bounce & $-3.39$ & $-6.29\times10^{-1}$ & $0$ & $-3.39$ & $-6.29\times10^{-1}$ & $0$ \\ 
		Turn & $-3.47$ & $0$ & $2.31\times10^{-1}$ & $-3.46$ & $0$ & $2.18\times10^{-1}$ \\ 
		Scaling & $-3.40$ & $6.04\times10^{-1}$ & $4.48\times10^{-1}$ & $-3.40$ & $4.95\times10^{-1}$ & $4.39\times10^{-1}$ \\
		KE = PE & $-0.55$ & $5.19\times10^{-6}$ & $4.70\times10^{4}$& $-0.55$ & $5.14\times10^{-6}$ & $4.70\times10^{4}$ \\
		End of Scaling & $-0.169$ & $1.10\times10^{-6}$ & $2.22\times10^{5}$ & $-0.169$ & $1.062\times10^{-6}$ & $2.25\times10^{5}$ \\
		KE=PE & $0.101$ & $2.08\times10^{-7}$ & $8.55\times10^{5}$ & $0.093$ & $1.93\times10^{-7}$ & $7.64\times10^{5}$ \\
		Turn & $0.141$ & $0$ & $1.27\times10^{6}$ & $0.127$ & $0$ & $1.15\times10^6$ \\
		$*$ & TFE & TFE & TFE & TFE & TFE & TFE \\
		\hline 
		ii. Bounce & $3.63$ & $-9.05\times10^{-1}$ & $0$ & $3.63$ & $-9.05\times10^{-1}$ & $0$ \\ 
		End of SI & $3.49$ & $-6.40\times10^{-1}$ & $1.80\times10^{-1}$ & $3.49$ & $-6.40\times10^{-1}$ & $1.80\times10^{-1}$ \\ 
		KE = PE & $1.20$ & $-6.09\times10^{-7}$ & $2.67\times10^{5}$ & $1.23$ & $-6.09\times10^{-7}$ & $2.36\times10^{5}$ \\
		Slow-Roll & $1.03$ & $-6.02\times10^{-9}$ & $1.97\times10^{6}$ & $1.08$ & $-4.97\times10^{-9}$ & $2.59\times10^{6}$ \\
		$*$ & TFE & TFE & TFE & $1.08$ & $-4.97\times10^{-9}$ & $2.58\times10^{7}$ \\
		\hline
		iii. Bounce & $3.67$ & $-9.05\times10^{-1}$ & $0$ & $3.67$ & $-9.05\times10^{-1}$ & $0$ \\ 
		End of SI & $3.53$ & $-6.40\times10^{-1}$ & $1.80\times10^{-1}$ & $3.53$ & $-6.40\times10^{-1}$ & $1.80\times10^{-1}$ \\ 
		KE = PE & $1.24$ & $-6.09\times10^{-7}$ & $2.67\times10^{5}$ & $1.27$ & $-6.10\times10^{-7}$ & $2.35\times10^{5}$ \\
		Slow-Roll & $1.07$ & $-5.09\times10^{-9}$ & $2.03\times10^{6}$ & $1.12$ & $-4.19\times10^{-9}$ & $2.64\times10^{6}$ \\
		$*$ & TFE & TFE & TFE & $1.08$ & $-4.89\times10^{-9}$ & $1.19\times10^{7}$ \\
		\hline
		iv. Bounce & $3.68$ & $-9.05\times10^{-1}$ & $0$ & $3.68$ & $-9.05\times10^{-1}$ & $0$ \\ 
		End of SI & $3.54$ & $-6.40\times10^{-1}$ & $1.80\times10^{-1}$ & $3.54$ & $-6.40\times10^{-1}$ & $1.80\times10^{-1}$ \\ 
		KE = PE & $1.25$ & $-6.10\times10^{-7}$ & $2.67\times10^{5}$ & $1.28$ & $-6.10\times10^{-7}$ & $2.35\times10^{5}$ \\
		Slow-Roll & $1.08$ & $-4.88\times10^{-9}$ & $2.03\times10^{6}$ & $1.13$ & $-4.02\times10^{-9}$ & $2.66\times10^{6}$ \\
		$*$ & $1.08$ & $-4.92\times10^{-9}$ & $2.90\times10^{6}$ & $1.08$ & $-4.89\times10^{-9}$ & $1.44\times10^{7}$ \\
		\hline
	\end{tabular}
	\caption{Comparison of Analytic (A) and Numerical (N) \cite{Bonga-Gupt} results for the evolution of observables for the case $\dot{\phi}_{B}<0$ for the Starobinsky potential. The last three blocks (ii-iv) are KE dominated cases wherein each row represents an event in ascending order: Bounce, End of Superinflation, Equilibrium, Turnaround, Slow-roll and Pivot point. The first block (i) is for the PE dominated case and contains more rows representing more events. In particular `Scaling' refers to when the solution Eqn.~(\ref{scaling-eqn}) holds and `End of Scaling' to when it breaks down. It does not lead to the desired slow-roll, which instead only occurs in the KE dominated regime with $\phi_{B}\geq 3.68m_{\rm Pl}$ for the analytic and $\phi_{B} \geq 3.63m_{\rm Pl}$ for the numerical cases. TFE means too few e-foldings}\label{Table4}
\end{center}
\end{table}

\subsection{$V(\phi)=\lambda\phi^{4}$} \label{Quartic-potential}
\noindent The basic behaviour of the inflaton field for a quartic potential is very similar to that of the quadratic potential (both for KE and PE dominated cases at the bounce). Thus using the procedure outlined in  \cite{Bonga-Gupt} we get the values of $\lambda$ in the potential and observables at the pivot point compatible with observations as: $\lambda=3.97\times10^{-14}$, $\phi_{*}=\pm 4.406m_{\rm Pl}$, $\dot{\phi}_{*}=\mp 4.03\times10^{-7}m_{\rm Pl}^{2}$ and $\phi_{end}=0.5642m_{\rm Pl}$. \\~\\
Tables~\ref{Table5} and \ref{Table6} describe the KE dominated cases at the bounce while the PE dominated cases are not included for the same reason as in the quadratic case i.e. the inflaton hardly moves from one point to another, hence the analytic and numerical results agree completely with one another, and usual slow roll inflation occurs. 
In Table~\ref{Table5}, consider the case $\phi_{B}=2.35m_{\rm Pl}$ with $\dot{\phi}_{B}\approx0.905m^2_{\rm Pl}$. Using Eqns.~(\ref{phidot-KEdom}-\ref{phi-KEdom}), at the end of the period of superinflation we determine $(\phi,\dot{\phi})=(2.49m_{\rm Pl},6.40\times10^{-1}m^2_{\rm Pl})$, matching exactly the numerical result.  Equating the kinetic and potential energies of the inflaton, the equilibrium point is then determined to be $(\phi_{eq},\dot{\phi}_{eq})=(4.42m_{\rm Pl},5.50\times10^{-6}m^2_{\rm Pl})$ with percentage errors $(0.7,1.3)$ compared to the numerical result. These equilibrium results serve as the initial conditions for the damping phase. Following the analysis of section~\ref{KEDom-Damping-Phase} we find the value of $\phi_{Turn}$ at the turnaround point to be $4.55m_{\rm Pl}$ compared to $\phi_{Turn}=4.50m_{\rm Pl}$ obtained numerically, again showing excellent agreement. Similar agreement emerges for the slow-roll point, $(\phi,\dot{\phi})=(4.52m_{\rm Pl},-4.10\times10^{-7}m^2_{\rm Pl})$ (analytical) compared to $(4.46m_{\rm Pl},-4.04\times10^{-7}m^2_{\rm Pl})$ (numerical). In this case, since $\phi_{SRA}>\phi_{*}=4.406m_{\rm Pl}$, it means that slow-roll is ocurring in a regime compatible with observations and we can use Eqn.~(\ref{phidot-SRA}) to determine the inflaton speed at the pivot point $\dot{\phi}=-\sqrt{\frac{2\lambda m^2_{\rm Pl}}{3\pi}}\phi_{*}=-4.04\times10^{-7}m^2_{\rm Pl}$, exactly matching the numerical result. \\~\\
As with the quadratic potential, the critical value of $\phi_B$ which ensures an observationally consistent period of slow roll differs slightly between the analytic and numerical cases. From Table~\ref{Table5} ($\phi_B >0$) we have $\phi_B \geq 2.23m_{\rm Pl}$ (analytically) and $\phi_B \geq 2.30m_{\rm Pl}$ (numerically). In Table~\ref{Table6} ($\phi_B <0$) we have $\phi_B \leq -6.67m_{\rm Pl}$ (analytically) and $\phi_B \leq -6.66m_{\rm Pl}$ (numerically). However, given the approximations that had to be made in obtaining the analytic solutions, the consistency of the two approaches is very encouraging. 

\begin {table}[H]
\begin{center}
	\begin{tabular}{ |c||c|c|c||c|c|c| } 
		\hline
		Event & $\phi_{A}/m_{\rm Pl}$ & $\dot{\phi_{A}}/m^2_{\rm Pl}$ & $t_{A} m_{\rm Pl}$ & $\phi_{N}/m_{\rm Pl}$ & $\dot{\phi_{N}}/m^2_{\rm Pl}$ & $t_{N}m_{\rm Pl}$ \\
		\hline
		i. Bounce & $2.20$ & $9.05\times10^{-1}$ & $0$ & $2.20$ & $9.05\times10^{-1}$ & $0$ \\ 
		End of SI & $2.34$ & $6.40\times10^{-1}$ & $1.80\times10^{-1}$ & $2.34$ & $6.40\times10^{-1}$ & $1.80\times10^{-1}$ \\ 
		KE = PE & $4.28$ & $5.16\times10^{-6}$ & $3.16\times10^{4}$ & $4.25$ & $5.09\times10^{-6}$ & $2.79\times10^{4}$ \\
		Turn & $4.41$ & $0$ & $1.13\times10^{5}$ & $4.36$ & $0$ & $1.04\times10^{5}$ \\
		Slow-Roll & $4.38$ & $-3.97\times10^{-7}$ & $2.00\times10^{5}$ & $4.32$ & $-3.91\times10^{-7}$ & $2.38\times10^{5}$  \\
		$*$ & TFE & TFE & TFE & TFE & TFE & TFE \\
		\hline
		ii. Bounce & $2.23$ & $9.05\times10^{-1}$ & $0$ & $2.23$ & $9.05\times10^{-1}$ & $0$ \\ 
		End of SI & $2.37$ & $6.40\times10^{-1}$ & $1.80\times10^{-1}$ & $2.37$ & $6.40\times10^{-1}$ & $1.80\times10^{-1}$ \\ 
		KE = PE & $4.31$ & $5.23\times10^{-6}$ & $3.11\times10^{4}$ & $4.28$ & $5.15\times10^{-6}$ & $2.75\times10^{4}$ \\
		Turn & $4.44$ & $0$ & $1.12\times10^{5}$ & $4.39$ & $0$ & $1.03\times10^{5}$ \\
		Slow-Roll & $4.41$ & $-4.00\times10^{-7}$ & $1.98\times10^{5}$ & $4.35$ & $-3.94\times10^{-7}$ & $2.35\times10^{5}$ \\
		$*$ & $4.406$ & $-4.04\times10^{-7}$ & $2.12\times10^{5}$ & TFE & TFE & TFE \\
		\hline
		iii. Bounce & $2.30$ & $9.05\times10^{-1}$ & $0$ & $2.30$ & $9.05\times10^{-1}$ & $0$ \\ 
		End of SI & $2.44$ & $6.40\times10^{-1}$ & $1.80\times10^{-1}$ & $2.44$ & $6.40\times10^{-1}$ & $1.80\times10^{-1}$ \\ 
		KE = PE & $4.37$ & $5.39\times10^{-6}$ & $3.02\times10^{4}$ & $4.34$ & $5.31\times10^{-6}$ & $2.67\times10^{4}$ \\
		Turn & $4.50$ & $0$ & $1.09\times10^{5}$ & $4.45$ & $0$ & $1.00\times10^{5}$ \\
		Slow-Roll & $4.48$ & $-4.06\times10^{-7}$ & $1.93\times10^{5}$ & $4.41$ & $-4.00\times10^{-7}$ & $2.89\times10^{5}$ \\
		$*$ & $4.406$ & $-4.04\times10^{-7}$ & $3.68\times10^{5}$ & $4.406$ & $-4.02\times10^{-7}$ & $2.47\times10^{5}$ \\
		\hline 
		iv. Bounce & $2.35$ & $9.05\times10^{-1}$ & $0$ & $2.35$ & $9.05\times10^{-1}$ & $0$ \\ 
		End of SI & $2.49$ & $6.40\times10^{-1}$ & $1.80\times10^{-1}$ & $2.49$ & $6.40\times10^{-1}$ & $1.80\times10^{-1}$ \\ 
		KE = PE & $4.42$ & $5.50\times10^{-6}$ & $2.96\times10^{4}$ & $4.39$ & $5.43\times10^{-6}$ & $2.61\times10^{4}$ \\
		Turn & $4.55$ & $0$ & $1.07\times10^{5}$ & $4.50$ & $0$ & $9.86\times10^{4}$ \\
		Slow-Roll & $4.52$ & $-4.10\times10^{-7}$ & $1.89\times10^{5}$ & $4.46$ & $-4.04\times10^{-7}$ & $2.24\times10^{5}$ \\
		$*$ & $4.406$ & $-4.04\times10^{-7}$ & $4.79\times10^{5}$ & $4.406$ & $-4.04\times10^{-7}$ & $3.58\times10^{5}$ \\
		\hline 
	\end{tabular}
	\caption{Comparison of Analytic (A) and Numerical (N) results for the evolution of observables for KE dominated cases with $\phi_{B}>0$ for a quartic potential, $V(\phi)=\lambda\phi^{4}$ where $\lambda=3.97\times10^{-14}$. Each block (i-iv) represents an initial condition at the bounce and each row within a block represents an event in ascending order: Bounce, End of Superinflation, Equilibrium, Turnaround, Slow-roll and Pivot point. An observationally consistent period of slow-roll happens when $\phi_{B}\geq 2.23m_{\rm Pl}$ for the analytic and $\phi_{B} \geq 2.30m_{\rm Pl}$ for the numerical simulations. TFE means too few e-foldings}\label{Table5}
\end{center}
\end{table}
\vspace*{-1.2cm}
\begin {table}[H]
\begin{center}
	\begin{tabular}{ |c||c|c|c||c|c|c| } 
		\hline
		Event & $\phi_{A}/m_{\rm Pl}$ & $\dot{\phi_{A}}/m^2_{\rm Pl}$ & $t_{A} m_{\rm Pl}$ & $\phi_{N}/m_{\rm Pl}$ & $\dot{\phi_{N}}/m^2_{\rm Pl}$ & $t_{N}m_{\rm Pl}$ \\
		\hline
		i. Bounce & $-6.60$ & $9.05\times10^{-1}$ & $0$ & $-6.60$ & $9.05\times10^{-1}$ & $0$ \\ 
		End of SI & $-6.46$ & $6.40\times10^{-1}$ & $1.80\times10^{-1}$ & $-6.46$ & $6.40\times10^{-1}$ & $1.80\times10^{-1}$ \\ 
		KE = PE & $-4.54$ & $5.81\times10^{-6}$ & $2.80\times10^{4}$ & $-4.57$ & $5.88\times10^{-6}$ & $2.47\times10^{4}$ \\
		Slow-Roll & $-4.33$ & $4.01\times10^{-7}$ & $1.40\times10^{5}$ & $-4.35$ & $4.02\times10^{-7}$ & $2.40\times10^{5}$  \\
		$*$ & TFE & TFE & TFE & TFE & TFE & TFE \\
		\hline
		ii. Bounce & $-6.66$ & $9.05\times10^{-1}$ & $0$ & $-6.66$ & $9.05\times10^{-1}$ & $0$ \\ 
		End of SI & $-6.52$ & $6.40\times10^{-1}$ & $1.80\times10^{-1}$ & $-6.52$ & $6.40\times10^{-1}$ & $1.80\times10^{-1}$ \\ 
		KE = PE & $-4.60$ & $5.97\times10^{-6}$ & $2.73\times10^{4}$ & $-4.63$ & $6.05\times10^{-6}$ & $2.40\times10^{4}$ \\
		Slow-Roll & $-4.40$ & $4.07\times10^{-7}$ & $1.37\times10^{5}$ & $-4.41$ & $4.08\times10^{-7}$ & $2.34\times10^{5}$ \\
		$*$ & TFE & TFE & TFE & $-4.406$ & $4.06\times10^{-7}$ & $2.54\times10^{5}$ \\
		\hline
		iii. Bounce & $-6.67$ & $9.05\times10^{-1}$ & $0$ & $-6.67$ & $9.05\times10^{-1}$ & $0$ \\ 
		End of SI & $-6.53$ & $6.40\times10^{-1}$ & $1.80\times10^{-1}$ & $-6.53$ & $6.40\times10^{-1}$ & $1.80\times10^{-1}$ \\ 
		KE = PE & $-4.61$ & $6.00\times10^{-6}$ & $2.71\times10^{4}$ & $-4.64$ & $6.08\times10^{-6}$ & $2.39\times10^{4}$ \\
		Slow-Roll & $-4.41$ & $4.08\times10^{-7}$ & $1.37\times10^{5}$ & $-4.42$ & $4.09\times10^{-7}$ & $2.33\times10^{5}$ \\
		$*$ & $-4.406$ & $4.04\times10^{-7}$ & $1.49\times10^{5}$ & $-4.406$ & $4.05\times10^{-7}$ & $2.80\times10^{5}$ \\
		\hline 
		iv. Bounce & $-6.70$ & $9.05\times10^{-1}$ & $0$ & $-6.70$ & $9.05\times10^{-1}$ & $0$ \\ 
		End of SI & $-6.56$ & $6.40\times10^{-1}$ & $1.80\times10^{-1}$ & $-6.56$ & $6.40\times10^{-1}$ & $1.80\times10^{-1}$ \\ 
		KE = PE & $-4.65$ & $6.09\times10^{-6}$ & $2.68\times10^{4}$ & $-4.68$ & $6.17\times10^{-6}$ & $2.36\times10^{4}$ \\
		Slow-Roll & $-4.44$ & $4.11\times10^{-7}$ & $1.35\times10^{5}$ & $-4.46$ & $4.12\times10^{-7}$ & $2.29\times10^{5}$ \\
		$*$ & $-4.406$ & $4.04\times10^{-7}$ & $2.27\times10^{5}$ & $-4.406$ & $4.04\times10^{-7}$ & $3.57\times10^{6}$ \\
		\hline 
	\end{tabular}
	\caption{Comparison of Analytic (A) and Numerical (N) results for the evolution of observables for KE dominated cases with $\phi_{B}<0$ for a quartic potential, $V(\phi)=\lambda\phi^{4}$ where $\lambda=3.97\times10^{-14}$. Each block (i-iv) represents an initial condition at the bounce and each row within a block represents an event in ascending order: Bounce, End of Superinflation, Equilibrium, Slow-roll and Pivot point. An observationally consistent period of slow-roll happens when $\phi_{B}\leq -6.67m_{\rm Pl}$ for the analytic and $\phi_{B} \leq -6.66m_{\rm Pl}$ for the numerical simulations. TFE means too few e-foldings}\label{Table6}
\end{center}
\end{table}

\section{Conclusion}
\label{conc}
\noindent In this paper we have revisited the interesting question of how does slow roll inflation emerge after the bounce in LQC? In particular we have sought to provide an accurate analytic treatment to complement and understand the numerical results presented for the case of a quadratic inflaton potential in \cite{Ashtekar-Sloan} and the Starobinsky potential in \cite{Bonga-Gupt}. We have also included the case of the quartic potential. Remarkably we can obtain excellent analytic agreement compared to the numerical solutions of the full non-linear evolution equations of LQC. Our approach has been to solve in different regimes and match solutions between those regimes. In particular we have divided the entire evolution of the inflaton from the initial bounce onwards into regimes dominated by different physical effects, allowing us to use suitable approximations in these regimes to obtain analytic solutions. For example in the case of the Starobinsky potential we were able to use the scaling behaviour of scalar fields in the presence of exponential potentials \cite{Copeland:1997et,Bahamonde:2017ize} and in braneworld settings \cite{Copeland:2004qe} to allow us to accurately solve for the evolution of the field showing that slow roll does not occur in the PE dominated regime when $\phi_{B}<0$. In all the cases studied our results can then be directly compared to the numerical results for the same initial bounce, and these are presented in Tables~\ref{Table1}-\ref{Table6}. The results are very encouraging, the bounds we obtain analytically on the critical value the inflaton field must have at the bounce for there to be a successful period of slow roll inflation matches the numerical values to within 6\%, and the subsequent evolution of $\phi$ and $\dot{\phi}$ are matched to within 5\% as can be seen in the tables. It is not obvious that this should have occurred given the different regimes of evolution the inflaton field experiences from super inflation just after the bounce, through to a damping regime as it climbs up the potential, a turn around followed by a slow roll regime. Our analysis works for both the cases where the fields energy is dominated at the bounce by either the Kinetic Energy of the field or its Potential Energy. For both the quadratic and quartic cases there are successful solutions in both regimes, whereas for the Starobinsky potential we find that if the field is initially dominated by its potential energy (with $\phi_B < 0$), then we find there is no successful slow roll period, whereas it can be found in the Kinetic Energy dominated regime as seen in Table~\ref{Table4}.  \\~\\
There are  a number of directions that we could take this analysis. The first is to consider the wide class of potentials available for the inflaton and see how they map onto the initial conditions provided by LQC, which in effect places a constraint on the allowed values of the field and its velocity as it emerges from the bounce (see for example \cite{Shahalam:2017wba}). A second would be to consider applying these techniques to the class of modified LQC models considered recently in Refs.~ \cite{Li:2018opr,Li:2018fco}. A third direction would be to see how these analytic solutions could help in the calculations of the density perturbations associated with the inflaton field. For example the Mukhanov-Sasaki equation applied to LQC can be written as \cite{Agullo-Ashtekar}  
\begin{eqnarray}
\ddot{Q}_{k}+3H\dot{Q}_{k}+\frac{k^{2}+U(t)}{a^{2}}Q_{k} & = & 0 \\
\ddot{T}_{k}+3H\dot{T}_{k}+\frac{k^{2}}{a^{2}}T_{k} & = & 0
\end{eqnarray}
where $U(t)=a^{2}\Big(f^{2}V(\phi)+2fV_{,\phi}(\phi)+V_{,\phi\phi}(\phi)\Big)$ is  a time dependent scalar potential and $f=\frac{\sqrt{24\pi}\dot{\phi}}{m_{\rm Pl} \sqrt{\rho}}$. Here the Mukhanov-Sasaki variables $Q_{k}(t)$ and $T_{k}(t)$ denote scalar and tensor perturbations in momentum space for a mode $k$ respectively. They are functions of the proper time. 
Given that $U(t)$ is a function of $\phi$ and $\dot{\phi}$, we are in a position to apply our analytic expressions for them, and in doing so in principle to obtain analytic solutions for $Q_{k}(t)$ and $T_{k}(t)$ applicable in the various regimes of validity of our solutions. The results can then be compared with the full numerical solutions, but with the clear advantage that comes hand in hand with having analytic solutions to fully understand the physics of a situation. One aim would be to obtain an analytical understanding for arbitrary potentials of the suppression in the power spectrum calculated on large scales which are seen numerically in the context of LQC. If this happens then we will have an analytic way of understanding how LQC leads to different results compared to standard inflation, and possibly of testing the paradigm. Similar work has been discussed in \cite{Zhu Old} and \cite{Zhu New}.

\acknowledgments
\noindent The authors are grateful to Abhay Ashtekar, Ivan Agullo, Beatrice Bonga, Brajesh Gupt, Jorge Pullin and especially to Parampreet Singh for useful comments. We are particularly grateful to the referee for their constructive comments, and to Steffen Gielen for very useful conversations. AB acknowledges financial support from the Department of Physics and Astronomy at LSU. EJC and JL acknowledge financial support from STFC consolidated grant No. ST/L000393/1 and ST/P000703/1. AB would like to acknowledge hospitality from the University of Nottingham where some of this work was completed during his M.Res.

\end{document}